\documentclass[aps,prb,reprint,groupedaddress,twocolumn,showkeys,amsmath,amssymb,longbibliography,floatfix]{revtex4-1}
\usepackage[english]{babel}
\usepackage{amsmath,amssymb,mathrsfs,bm,braket,color,graphicx,comment,amsfonts,mwe,tikz}
\usepackage{tabularx}
\usepackage[export]{adjustbox}
\usepackage[percent]{overpic}

\usepackage{contour,color,stackengine,titletoc,lipsum,tikz}
\usepackage[colorlinks,citecolor=blue,urlcolor=blue]{hyperref}
\usepackage[mathscr]{euscript}	
\usepackage[normalem]{ulem}

\newcommand{\vect}[1]{\boldsymbol{#1}}

\newcommand{\coefs}{\eta}
\newcommand{\PLL}{\mathcal{P}_{\text{\tiny LLL}}}

\newcommand\thefont{\expandafter\string\the\font}

\begin{document}

\title{Large-scale simulations of particle-hole-symmetric Pfaffian trial wave functions}

\author{Mykhailo Yutushui}
\author{David F. Mross}
\affiliation{Department of Condensed Matter Physics, Weizmann Institute of Science, Rehovot, 76100, Israel}

\date{\today}
\begin{abstract}
We introduce a family of paired-composite-fermion trial wave functions for any odd Cooper-pair angular momentum. These wave functions are parameter-free and can be efficiently projected into the lowest Landau level. We use large-scale Monte Carlo simulations to study three cases: Firstly, the Moore-Read phase, which serves us as a benchmark. Secondly, we explore the pairing associated with the anti-Pfaffian and the particle-hole-symmetric Pfaffian. Specifically, we assess whether their trial states feature exponentially decaying correlations and thus represent gapped phases of matter. For Moore-Read and anti-Pfaffian we find decay lengths of $\xi_\text{Moore-Read}=1.30(5)$ and $\xi_\text{anti-Pfaffian}=1.38(14)$, in units of the magnetic length. By contrast, for the case of PH-Pfaffian, we find no evidence of a finite length scale for up to $56$ particles.
\end{abstract}
\maketitle

\section{Introduction}

The half-filled Landau level is well known for realizing two particularly remarkable manifestations of strong electronic correlations. In the lowest Landau level (LLL) at filling factor $\nu=\frac{1}{2}$, Coulomb interactions between electrons result in a metallic state of emergent charge-neutral quasiparticles.\cite{willett_experimental_1993,Kang_how_1993,Goldman_detection_1994,Rezayi_fermi-liquid-like_1994,Smet_magnetic_1996}
The properties of this state are well described in terms of composite fermions (CFs)---electrons bound to two fictitious flux quanta.~\cite{Jain_composite-fermion_1989,Jain_quantitative_1997,Jain_composite_2007} At half-filling, these flux quanta compensate for the magnetic field, and CFs form a gapless ``composite Fermi liquid'' (CFL).\cite{halperin_theory_1993} In the half-filled first excited Landau level, i.e., at $\nu=\frac{5}{2}$, there is instead strong numerical and experimental evidence of a gapped state\cite{Willett_observation_1987,Morf_transition_1998,Rezayi_incompressible_2000,Wojs_landau_level_2010,Storni_fractional_2010,feiguin_density_2008,Feiguin_spin_2009,Peterson_Finite_Layer_Thickness_2008,Rezayi_breaking_2011,Pakrouski_phase_diagram_2015} attributed to pairing of composite fermions.~\cite{Halperin_QH_1983,Moore_nonabelions_1991,Greiter_half_filled_1991,Haldane_spin_singlet_1988,Read_paired_2000} The pairing channel is still not fully settled; the leading contenders in numerical studies are the Moore-Read Pfaffian state, which features Cooper pairs with angular momentum $\ell=-1$,~\cite{Moore_nonabelions_1991,Read_paired_2000} and the anti-Pfaffian with $\ell=3$.~\cite{Storni_fractional_2010,Levin_particle-hole_2007,Lee_particle-hole_2007} 

A subtle issue in studies of the half-filled Landau level is particle-hole (PH) symmetry, which arises when electrons occupy exactly half of a particular Landau level and interact solely via two-body interactions. Analyses of PH symmetry date back several decades,~\cite{Girvin_PHS_1984} but recent developments have refocused theoretical and experimental efforts on this question. The first of these was a proposal to describe CFs as Dirac particles, on which PH symmetry acts as time-reversal symmetry.~\cite{Son_is_2015,Metlitski_particle_vortex_2016,Wang_half_filled_2016,Geraedts_half_filled_2016,Murthy_half_filled_2016,Kachru_half_filled_2016,Mross_explicit_duality_2016,Mulligan_emergent_ph_2016,Balram_nature_2016,Yang_dirac_2017,Fremling_trial_2018}
The pairing of Dirac CFs is related to one of non-relativistic CFs via $\ell_\text{Dirac} = 1 -\ell$. In particular, the unique pairing channel consistent with PH symmetry has $\ell_\text{Dirac}=0$ or $\ell=1$. The corresponding topological order, irrespective of whether this symmetry is preserved or not, is known as ``PH-Pfaffian''.\cite{Son_is_2015,chen_symmetry_2014,Bonderson_time-reversal_2013,Metlitski_symmetry_respecting_2015,Wang_gapped_2013,Mross_Composite_Dirac_Liquids_2015} To date, there is no known microscopic model that realizes this phase. For the case of Coulomb-interacting electrons at $\nu=\frac{5}{2}$ without Landau-level mixing,\cite{Morf_transition_1998} the numerical findings imply a spontaneously broken PH symmetry. 

Experimentally, distinguishing between the possible paired states at $\nu=\frac{5}{2}$ is notoriously difficult because they exhibit the same universal responses to all charge-sensitive probes. In particular, the Hall conductance $\sigma_{xy}=\frac{5}{2}\frac{e^2}{h}$ is independent of $\ell$, as is the elementary quasiparticle charge $e^*=\frac{e}{4}$, measured in Ref.~\onlinecite{Dolev_observation_2008}. It has long been appreciated that measurements of the thermal Hall conductance $\kappa_{xy}$ can distinguish between these states, with $\kappa_{xy}=(3-\frac{\ell}{2}) \frac{\pi k_\text{B}}{3 h}T$. Such an experiment was recently carried out and, surprisingly, found the value $\ell=1$ corresponding to PH-Pfaffian.~\cite{Banerjee_observation_2018} The marked absence of this phase from all numerical studies has led to proposals that the observed value is either due to disorder-induced formation of the PH-Pfaffian topological order,~\cite{Mross_theory_2018,Lian_theory_2018,Wang_topological_2018} or reflects incomplete equilibration between the edge states of an underlying anti-Pfaffian phase.~\cite{Simon_equilibration_2018,Feldman_comment_2018,Simon_reply_comment_2018,Feldman_equilibration_2019,Simon_equilibration_2020,Asasi_equilibration_2020}

The numerical search for a realization of the PH-Pfaffian phase in a clean quantum Hall system is considerably impeded by the absence of a trial wave function. A generalization of the celebrated Moore-Read wave function to the PH-Pfaffian phase was proposed in Refs.~\onlinecite{Zucker_stabilization_2016,Yang_particle-hole_2017} (a related wave function was previously introduced in Ref.~\onlinecite{Jolicoeur_new_series_2007}). However, subsequent numerical studies of this state have raised doubts whether it describes a gapped phase or instead represents a gapless CFL.\cite{Balram_parton_2018,Mishmash_numerical_2018}

Specifically, these works observed a substantial overlap between PH-Pfaffian and composite Fermi liquid trial states, as well as pronounced $2k_{\text{F}}$ oscillations in the density-density correlation function. (Within BCS mean-field theory, the amplitude of such oscillations decays exponentially with a length scale set by the inverse superconducting gap.) These analyses were, however, limited to relatively small systems of $N_\text{p}=12$ particles by the need to perform an explicit projection into the LLL. In such small systems, the density-density correlation function exhibits less than two full oscillations. Finite-size effects are thus dominant and the asymptotic decay cannot be determined. In this work, we develop a method of studying PH-Pfaffian trial wave functions for much larger particle numbers and present numerical results for up to $N_\text{p}=56$ particles.

The rest of this paper is organized as follows: In Sec.~\ref{sec.pairedstates}, we introduce a family of paired-composite-fermion wave functions for arbitrary odd $\ell$, which can be efficiently studied via Monte Carlo methods. This family includes, in particular, the Moore Read ($\ell=-1$), the anti-Pfaffian ($\ell=3$), and the PH-Pfaffian ($\ell=1$)  states. In Sec.~\ref{sec.LLL}, we describe and compare several different routes of obtaining LLL wave function from composite-fermion based \textit{Ans\"atze}. Sec.~\ref{sec.num} contains our main numerical results. 
We compute density-density correlation functions of all three states as well as those of unpaired CFLs. We also calculate the overlap between paired and unpaired trial states. In Sec.~\ref{sec.conc}, we conclude by discussing the implications of our findings and potential directions in the search for suitable PH-Pfaffian trial states. The Appendices contain additional numerical data and technical details regarding LLL projection.

\section{Trial wave functions of paired-composite fermions}\label{sec.pairedstates}
A wide class of paired-composite-fermions wave functions on a sphere was introduced in Refs.~\onlinecite{moller_paired_2008,Moller_phd} as
\begin{align}
\label{eqn.MS}
 \Psi_\text{M\"oller-Simon} =\PLL\left\{ \text{Pf}\left[g_{ab}\right] \prod\nolimits_{a<b}\omega_{ab}^{2p}\right\}~,
\end{align}
where $\omega_{ab}= u_a v_b - u_b v_a$ with $u = \cos(\frac{\theta}{2})\exp(i\frac{\phi}{2})$ and $v=\sin(\frac{\theta}{2})\exp(i\frac{\phi}{2})$. In general, these wave functions require an explicit projection into the LLL, denoted by $\PLL$. The Pfaffian factor alone describes a BCS superconductor with Cooper-pair wave function $g_{ab}$ and the Jastrow factor $\prod\nolimits_{a<b}\omega_{ab}^{2p}$ encodes attachment of $2p$ flux quanta. Instead of the magnetic flux $N_\phi$, CFs experience the reduced flux $N_\phi^\text{eff} =N_\phi -2p(N_\text{p}-1) $. Due to the curvature of the sphere, the particle number may be offset from the product of filling factor and magnetic flux by the shift $\mathcal{S}= \nu^{-1}N_\text{p} -N_\phi$.~\cite{Wen_shift_1992} In the half-filled LLL, the effective monopole strength is determined by the shift according to $q \equiv N_\phi^\text{eff}/2=1 - \mathcal{S}/2$.

In order to specify the pairing channel, we expand $g_{ab}$ in single-particle orbitals corresponding to $q$, i.e., monopole harmonics ${ Y}^q_{l,m}$. \footnote{Monopole harmonics and their properties were described in Refs.~\onlinecite{Wu_dirac_1976,Wu_properties_1977}; we use the conventions of Ref.~\onlinecite{Jain_composite_2007}} (Ref.~\onlinecite{moller_paired_2008} focused on the Moore-Read case $q_\text{MR} = -\frac{1}{2}$. In the context of bilayer systems the case $q = \frac{1}{2}$ was studied in Refs.~\onlinecite{moller_paired_2008,Moller_Trial_2009}.) In states with a well-defined Cooper-pair angular momentum $\ell=2q$, the monopole harmonics enter $g^\ell_{ab}$ through the linear combinations 
\begin{align}
\label{eqn.Phi}
\Phi^q_{l}\equiv \sum\nolimits_{m}{ Y}^{q}_{l,m}(r_a) {\overline{ Y}}^{q}_{l,m}(r_b) = \omega_{ab}^{2q}f^{q}_l(|\omega_{ab}|)~, 
\end{align}
where $ {\overline{ Y}}^{q}_{l,m}\equiv \bigl({{ Y}}^{-q}_{l,m}\bigr)^*$ and $f_{l}^q$ are specified in App.~\ref{app.addition}. (We use $\ell$ to denote the Cooper-pair angular momentum and $l$ for single-particle orbitals.) A general rotationally symmetric wave function can thus be parametrized as
\begin{align}
\label{eqn.g}
g^\ell_{ ab} = \sum\nolimits_{l\geq |q|} \coefs_l \Phi_l^q~.
\end{align}
Different choices of $\eta_l$ can realize both unpaired and paired states. In the latter cases, the particular choice of $\coefs_l$ does not affect the Cooper-pair angular momentum.

In general, $g_{ab}^\ell$ contains contributions of electrons in higher Landau levels and, therefore, an explicit LLL projection is required. The contributions from $\coefs_l$ with large $l$ vanish upon projection.\cite{moller_paired_2008} Consequently, the {\it a priori} infinite number of variational parameters becomes finite and $N_\text{p}$ dependent. The remaining $\coefs_l$ can be used to, e.g., optimize the energy of a trial state with respect to a given interaction potential. Conversely, to study specific phases, one needs a prescription for choosing the parameters $\coefs_l$ for any number of particles.

We propose the family of Cooper-pair wave functions,
\begin{align} \label{eqn.pair_WF}
 g^{\ell}_{ab}\equiv\frac{1}{|\omega_{ab}|}e^{i \ell \vartheta_{ab}}= \frac{\omega_{ab}^{q-1/2}}{\bar{\omega}_{ab}^{q+1/2}}~,
\end{align}
where $\vartheta_{ab} = \text{arg}(\omega_{ab})$. The pairing functions $g^{\ell}_{ab}$ have angular momentum $\ell$ and the characteristic weak-pairing decay $|g^\ell_{ab}|\propto \frac{1}{R}$ at long distances, see also App.~\ref{app.SC_WF}. [In the absence of any length scale, these two properties alone fix the Cooper-pair wave functions to be given by Eq.~\eqref{eqn.pair_WF}.] They can be brought into the form of Eq.~\eqref{eqn.g} through 
\begin{align} \frac{\omega_{ab}^{q-1/2}}{\bar{\omega}_{ab}^{q+1/2}}=\sum_{l,m}\dfrac{8\pi}{2l+1}{ Y}^q_{l,m}(r_a)\overline{ Y}^{q}_{l,m}(r_b)~.
 \label{eqn.addition}
\end{align}
For $q=0$, this identity reduces to the expansion of the Coulomb potential in spherical harmonics. The $q\neq 0$ case is its generalization to monopole harmonics. We have not seen this expression reported in the literature, but its derivation is straightforward (see App.~\ref{app.addition}).

\subsection{Composite Fermi liquid}
The wave function $\Psi_\text{M\"oller-Simon}$ of
Eq.~\eqref{eqn.MS} can describe CFL trial states~\cite{Rezayi_fermi-liquid-like_1994} for particle numbers that satisfy 
\begin{align}N_\text{p} =\sum\nolimits_{l = |q|}^{l_{\text{F}}} (2 l +1)=(l_\text{F} +1)^2- q^2.\label{filledshells}\end{align}
 If $\coefs_{l}=0$ for $l>l_\text{F}$, the Pfaffian contains precisely as many linearly independent orbitals as particles. Consequently, 
 \begin{subequations}
 \label{cflmain}
\begin{align}
\label{eqn.cfl}
 \Psi_\text{CFL}^{q}&\equiv
\PLL\left\{ \text{Pf}\left[\sum\nolimits_{l =|q|}^{l_{\text{F}} }\coefs_l \Phi_l\right] \prod\nolimits_{a<b}\omega_{ab}^2\right\}~,\\
& \propto\PLL\left\{ \det\left[Y^q_{l_a,m_a}(r_b)\right] \prod\nolimits_{a<b}\omega_{ab}^2\right\}~.
\label{eqn.cfl2}
\end{align}
 \end{subequations}
\normalsize
 Suppressing $\eta_l$ with $l>l_\text{F}$ forces composite fermions to occupy only states below $l_\text{F}$, i.e., to form a filled Fermi sea. The parameters $\coefs_l$ can thus be used to interpolate between unpaired and paired states.

\subsection{Moore-Read Pfaffian}\label{sec.wf.mr}
For the monopole strength $q_\text{MR}=-\frac{1}{2}$, the left-hand side of Eq.~\eqref{eqn.addition} is $1/\omega_{ab}$ and $\Psi_\text{M\"oller-Simon}$ reduces to the celebrated Moore-Read wave function~\cite{Moore_nonabelions_1991}
\begin{align}
\Psi_\text{Moore-Read} =\PLL\left\{ \text{Pf}\left[\frac{1}{\omega_{ab}}\right] \prod\nolimits_{a<b}\omega_{ab}^2\right\}~.
\label{eqn.MR}
\end{align}
This wave function is already in the LLL, and projection acts trivially. This property is no longer manifest when the argument of the Pfaffian is expanded using Eq.~\eqref{eqn.addition}; it only holds for the infinite sum. However, the contributions from $l > 2N_\text{p}-3$ vanish upon projection. It follows that $\Psi_\text{Moore-Read}$ can be reproduced \textit{exactly} with a \textit{finite} number of parameters $\coefs_l = \frac{8\pi}{2l+1}$.

Ref.~\onlinecite{moller_paired_2008} treated $\coefs_l$ as variational parameters and reported a numerically obtained overlap of over $99\%$ between the Moore-Read and the variational state for $N_\text{p}=20$. That finding was based on a modified form of projection into the lowest Landau level that spoils the exact agreement (see Sec.~\ref{sec.cf.single.projection} for details). A different implementation of the projection results in the exact Moore-Read wave function (see Sec.~\ref{sec.cf.pair.projection}).

\subsection{Anti-Pfaffian}\label{sec.wf.apf}
For monopole strength $q_\text{aPf}=\frac{3}{2}$, Eq.~\eqref{eqn.pair_WF} describes pairing with angular momentum $\ell=3$. Composite fermions with this pairing symmetry form the anti-Pfaffian phase. A trial state widely used in exact diagonalization studies is obtained through particle-hole conjugation of the Moore-Read wave function $\Psi_\text{anti-Pfaffian}\equiv \hat{\Theta}\Psi_\text{Moore-Read}$. This wave function is, however, not well suited for large-scale Monte Carlo simulations.~\cite{Wang_MC_PH_conjugation_2019} Recently, an elaborate parton construction was used to produce an alternative wave function, $\Psi^\text{parton}_\text{anti-Pfaffian}$, amenable to Monte Carlo methods.~\cite{Balram_parton_2018} For $N_\text{p} = 10$ it exhibits significant overlap $93.8\%$ with $\Psi_\text{anti-Pfaffian}$ and the same degeneracies of low-lying states in the entanglement spectrum. 

Based on the general family of wave functions defined through Eq.~\eqref{eqn.pair_WF}, we propose 
\begin{align}
\Psi^\text{CF}_\text{anti-Pfaffian} =\PLL\left\{ \text{Pf}\left[\frac{\omega_{ab}}{\bar \omega_{ab}^2}\right] \prod\nolimits_{a<b}\omega_{ab}^2\right\}~, 
\label{eqn.APF}
\end{align}
as an alternative trial state in the anti-Pfaffian phase. In contrast to the Moore-Read case, an explicit projection into the LLL is required here. All efficient algorithms for this purpose that have been developed over the years require that $\bar u$ and $\bar v$ appear with positive powers,~\cite{Jain_quantitative_1997,Davenport_projection_2012,Fulsebakke_projection_2016,Mukherjee_incompressible_2015} which is not the case in Eq.~\eqref{eqn.APF}. To simulate and analyze $\Psi^\text{CF}_\text{anti-Pfaffian}$ for large particle numbers, we therefore use the expansion of Eq.~\eqref{eqn.addition} with $q_\text{aPf}=\frac{3}{2}$. The resulting wave function contains only positive powers of $\bar u$ and $\bar v$, more specifically in the form of monopole harmonics. The wave function is thus in a form that can be efficiently projected into the lowest Landau level.

\subsection{Particle-hole-symmetric Pfaffian}\label{sec.wf.ph}
For monopole strength $q_\text{PH}=\frac{1}{2}$, Eq.~\eqref{eqn.pair_WF} describes pairing with angular momentum $\ell=1$. The corresponding paired-composite-fermion wave function is given by
\begin{align}
\Psi_\text{PH-Pfaffian} =\PLL\left\{ \text{Pf}\left[\frac{1}{\bar \omega_{ab}}\right] \prod\nolimits_{a<b}\omega_{ab}^2\right\}~.
\label{eqn.PH}
\end{align}
It was previously proposed in Ref.~\onlinecite{Zucker_stabilization_2016} and studied numerically for $N_\text{p}=10,12$ in Refs.~\onlinecite{Balram_parton_2018,Mishmash_numerical_2018}. These works found indications that projection may result in a state with surprisingly weak (or altogether absent) pairing between composite fermions. Specifically, the overlap between $\Psi_\text{PH-Pfaffian}$ and a CFL at the same shift is surprisingly large. 
For $N_\text{p}=12$ 
the overlap $|\langle \Psi_\text{PH-Pfaffian}|\text{CFL}\rangle| = 95.42(1)\%$ is significantly larger than the one between the Moore-Read state and CFL, $|\langle \Psi_\text{MR-Pfaffian}|\text{CFL}\rangle| =61.9(3)\%$, despite a substantially smaller Hilbert space of the latter.\cite{Mishmash_numerical_2018} Moreover, the density-density correlation function of $\Psi_\text{PH-Pfaffian}$ features more pronounced $2k_{\text{F}}$ oscillations than $\Psi_\text{Moore-Read}$. 

As for the case of anti-Pfaffian, the expansion of Eq.~\eqref{eqn.addition} provides us with a wave function that can be efficiently projected into the LLL.

\section{Lowest Landau-level projection}\label{sec.LLL}

Most of the trial states discussed above do not reside entirely in the LLL. To eliminate contributions from higher Landau levels, one may expand a wave function in single-particle orbitals with well-defined Landau-level index $n$, and retain only those with $n=0$. This form of projection was applied to PH-Pfaffian trial states in Refs.~\onlinecite{Balram_parton_2018,Mishmash_numerical_2018}, but the exponential increase of the Hilbert-space size quickly renders this approach unfeasible. The same LLL wave function can be obtained by replacing all instances of $\bar u$ and $\bar v$ with derivatives $\partial_u$ and $\partial_v$.~\cite{Jain_composite_2007} However, the number of required derivatives grows rapidly and becomes intractable with modern mathematical software for moderate $N_\text{p}\gtrsim 10$. To study larger systems, alternative routes for obtaining LLL wave functions from a given composite-Fermion ansatz have thus been developed.

\subsection{Single-composite-fermion projection}\label{sec.cf.single.projection}

The most widely used projection method was introduced in Ref.~\onlinecite{Jain_quantitative_1997} based on the ``composite-fermion orbitals''
\begin{align}
 {\cal Y}^{q}_{l,m}(r_a)=
 { Y}^{q}_{l,m}(r_a)\prod\nolimits_{b \neq a}\omega^p_{ab}~,
 \label{eqn.cforbital}
\end{align}
with $p=1$ at half-filling. The Pfaffian and Jastrow factor of Eq.~\eqref{eqn.MS} can be combined to write the wave function succinctly as
\begin{align}\label{eqn.CF_pair_form}
&\text{Pf}\left[g_{ab}\right] \prod\limits_{a < b}\omega_{ab}^2=
\text{Pf}\left[\sum\nolimits_{l,m}\coefs_l
 {\cal Y}^{q}_{l,m}(r_a){ \overline{\cal Y}}^{q}_{l,m}(r_b)\right]~.
\end{align} 
Here, $\overline {\cal Y}$ is defined as complex conjugation of the monopole harmonic, but not of the $\omega^p$ factor. 

One can now define ``single-composite-fermion projection'' as separately projecting each CF orbital, i.e.,
\small
\begin{align}\label{eqn.single}
&\PLL^\text{single} \text{Pf}[
\cdots]\equiv
\text{Pf}\left[
\sum_{l,m}\coefs_l
 \PLL{ \mathcal{Y}^{q}_{l,m}(r_a)}\PLL{ \overline{\cal Y}}^{q}_{l,m}(r_b)\right].
\end{align}
\normalsize

Algorithms for the efficient evaluation of the projected composite-fermion orbitals were described in Ref.~\onlinecite{Jain_quantitative_1997} and further refined in Refs.~\onlinecite{Davenport_projection_2012,Fulsebakke_projection_2016,Mukherjee_incompressible_2015}.

\begin{table}
\caption{
Single and pairwise projections of Eq.~\eqref{eqn.MR_exp} for $N_\text{p}=56$ are well converged at cutoff $N_\text{c}\equiv l_\text{c} - |q|=14$. The latter becomes indistinguishable from $\Psi_\text{Moore-Read}$ within double precision for $N_\text{c}\approx 18$ and matches exactly for $N_\text{c} \geq 55$.
}
\setlength{\extrarowheight}{2pt}
 \begin{tabularx}{\linewidth}{l |@{\hspace{2mm}} l l l l}
\hline\hline
 & $N_\text{c}=8$ & $N_\text{c}=10$ & $N_\text{c} =15$ & $N_\text{c} = 20$ \\[0.5mm]
\hline
$|\langle\Psi_{\text{MR}}|\Psi^{\text{pair}}_{\text{MR},l_\text{c}}\rangle | $ 
& $95.4(3)\%$ & $98.48(6)\%$ & $99.995(1)\%$ & $100(0)\%$ \\
$|\langle\Psi_{\text{MR}}|\Psi^{\text{single}}_{\text{MR},l_\text{c}}\rangle| $
& $77.2(6)\%$ & $95.4(2)\%$ & $93.3(2)\%$ & $93.3(2)\%$\\
\hline\hline
\end{tabularx}
\label{tab.convergence_MR}
\end{table}

\subsection{Pairwise composite fermion projection} \label{sec.cf.pair.projection}
The form of the wave function in Eq.~\eqref{eqn.CF_pair_form} suggests projecting the argument of the Pfaffian as a whole, rather than individual ${\cal Y}$. We thus define ``pairwise-composite-fermion projection'' as
\small
\begin{align}\label{eqn.pair}
&\PLL^\text{pair}\text{Pf}[
\cdots]\equiv
\text{Pf}\left[\sum\nolimits_{l,m}\coefs_l
\PLL\left\{{\cal Y}^{q}_{l,m}(r_a)
{\overline{\cal Y}}^{q}_{l,m}(r_b)\right\}\right]~.
\end{align}
\normalsize
This form of projection may be implemented similarly to single CF projection and with comparable efficiency. (When using the algorithms described in Refs.~\onlinecite{Fulsebakke_projection_2016,Mukherjee_incompressible_2015}, we find pairwise projection to be moderately faster.) In App.~\ref{app.Pairwise_projection}, we describe this approach in detail.

\subsection{ Comparison of different projection methods}

Wave functions projected with either $\PLL$, $\PLL^\text{single}$, or $\PLL^\text{pair}$ do not, in general, coincide. However, in most cases, they describe the same topological phase, which can be inferred, e.g., from their entanglement spectra. It is thus often justified to adopt single or pairwise projection to access large system sizes.

To illustrate the different projection methods, consider the (unprojected) wave function
\begin{align}\label{eqn.MR_exp}
 \Psi^{q}_{l_\text{c}} =\text{Pf}\left[\sum_{l=|q|}^{l_\text{c}}\frac{8\pi}{2l+1}\sum_{m}
 \mathcal{Y}^{q}_{l,m}(r_a)\overline{\mathcal{Y}}^{q}_{l,m}(r_b)\right]~.
\end{align}
For $q_\text{MR}$ and cutoff $l_\text{c} = \infty$\footnote{Upon rewriting Eq.~\eqref{eqn.addition} in the form of Eq.~\eqref{eqn.A_expanssion1}, it becomes manifest that the pairing channel is unaffected by truncating the infinite sum.}, it coincides with $\Psi_\text{Moore-Read}$, and the argument of the Pfaffian lies in the LLL. Consequently, both $\PLL$ and $\PLL^\text{pair}$ act trivially. At finite $l_\text{c}$, the argument contains contributions from higher Landau levels. Still, for $l_\text{c} > 2N_\text{p}-3$ and $l_\text{c} > N_\text{p}-1$, the Moore-Read wave function is reproduced exactly by $\PLL$ and $\PLL^\text{pair}$. Contributions from larger $l$ vanish under projection.\cite{Jain_quantitative_1997} By contrast, single-composite-fermion projection applies to each orbital separately and thus acts non-trivially for any $l_\text{c}$. In practice, both single and pairwise projections well approximate $\Psi_\text{Moore-Read}$ already at $l_\text{c} \approx 2 l_\text{F} \approx 2\sqrt{N_\text{p}}$, (see Tab.~\ref{tab.convergence_MR}). 

For $q_\text{aPf}$ and $q_\text{PH}$, we find that $\PLL^\text{single}$ and $\PLL^\text{pair}$ yield wave functions that are almost independent of $l_\text{c}$ beyond $2l_\text{F}$ (see App.~\ref{app.PH_supp} for data at $q_\text{PH}$). This rapid convergence makes the expansion attractive for numerical simulations. 

There are, however, instances where different types of projection yields dramatically different results. A striking example is given by the CFL at $q_\text{MR}$, which arises for $l_\text{c} = l_\text{F}$ [see also Eq.~\eqref{cflmain}]. Here, $\PLL$ and $\PLL^\text{single}$ produce the expected gapless states; for $N_\text{p}=12$, their overlaps with $\Psi_\text{Moore-Read}$ are $61.9(3)\%$ and $57.5(1)\%$, respectively. When using pairwise projection, we instead find a state with a much larger overlap of $97.9(1)\%$ and an entanglement spectrum matching that of $\Psi_\text{Moore-Read}$. For larger $N_\text{p}$, the margin between two projection schemes grows rapidly (see Table~\ref{tab.CFL}). By contrast, for CFLs at positive $q$, single and pairwise projections are identical. We therefore use single-composite-fermion projection (the Jain-Kamilla method) for CFLs at any $q$.

The example of the CFL at $q_\text{MR}$ illustrates that wave functions obtained through different projection schemes need not even belong to the same phase. In cases such as PH-Pfaffian, where projection yields unexpected results,\cite{Balram_parton_2018,Mishmash_numerical_2018} it may be prudent to compare different methods. Fortunately, this is \textit{not} the ``typical'' behavior, and in the case of PH-Pfaffian, we do not find significant differences between trial states projected with either method.
\begin{table}
\caption{
\label{tab.CFL}Single and pairwise projections of CFLs at $q_\text{MR}$ yield strikingly different overlaps with $\Psi_\text{Moore-Read}$. At $q_\text{PH}$, the overlap between CFL and $\Psi_\text{PH}$ is substantial, independent of the projection method. 
}
\setlength{\extrarowheight}{2pt}
 \begin{tabularx}{\linewidth}{c | >{\raggedleft\arraybackslash}X >{\raggedleft\arraybackslash}X >{\raggedleft\arraybackslash}X >{\raggedleft\arraybackslash}X}
\hline\hline
 & $N_\text{p} = 12$ \ & $N_\text{p} = 30$\ \ & $N_\text{p} = 42$\ \ \ & $N_\text{p} = 56$\ \ \ \\
\hline
$|\langle \Psi^{\text{pair}}_{\text{PH}}|\Psi_{\text{CFL}}^\text{single}\rangle|$ & $95.8(1)\%$ & $66.7(1)\%$ & $ 46.2(1)\%$ & $28.7(1)\%$\\
$|\langle \Psi_{\text{MR}}|\Psi^{\text{single}}_{\text{CFL}}\rangle|$ & $57.5(1)\%$ & $ 13.2(2)\%$ & $4.8(5)\%$ & $1.3(1)\%$\\
$|\langle \Psi_{\text{MR}}|\Psi^{\text{pair}}_{\text{CFL}}\rangle|$ & $97.9(1)\%$ & $ 94.4(2)\%$ & $ 91.9(4)\%$ & $89.6(4)\%$\\
\hline\hline
\end{tabularx}
\end{table}
\section{Numerical results}\label{sec.num}
\begin{figure*}
\includegraphics[width=\textwidth ,frame]{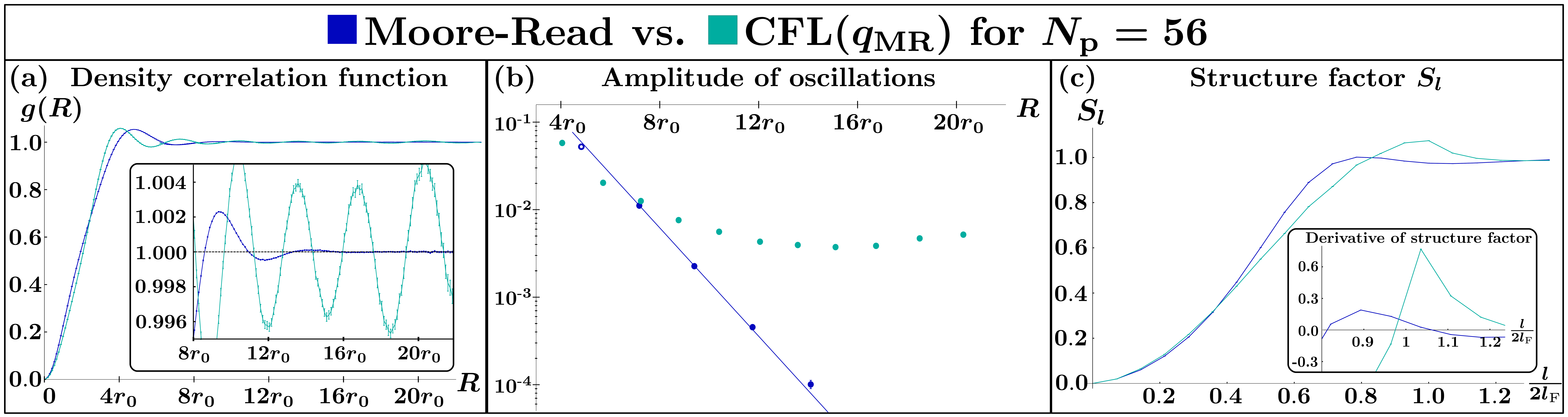}
\caption{
(a) The density-density correlation function of the Moore-Read state exhibits $2k_{\text{F}}$ oscillations that decay very rapidly with the arc length $R$. In the CFL, by contrast, oscillations persist across the entire sphere. (b) A semilogarithmic plot of the oscillation amplitude shows their exponential decay in the Moore-Read case. The straight line represents a fit to a decay length $\xi \approx 1.46r_0$, where the maximum near the typical inter-particle distance $R= 4r_0$ is viewed as encoding short-distance properties and thus excluded. The amplitude for the CFL decays much slower than exponentially. (c) The static structure factor of the Moore-Read state is smooth for all $l$, while that of the CFL exhibits a sharp cusp at $2l_{\text{F}}$, indicating a power-law decay of real-space oscillations.\label{fig.cor_mr_cfl}}
\end{figure*}
The primary motivation for this work is to determine whether the PH-Pfaffian wave function represents a gapped phase. Since we are studying properties of trial states rather than Hamiltonians, we use a finite correlation length $\xi$ as a proxy for the (inverse) gap. 
 Using Monte Carlo simulation, we compute the normalized density-density correlation function on a unit sphere 
\begin{align}\label{eqn.paircorr}
 g(R) = \dfrac{1}{N^2_p}\sum\nolimits_{a\neq b} \left\langle \delta\left[\vect r_a\cdot \vect r_b - \cos(R)\right] \right\rangle~,
\end{align} where $\vect r_{a,b}$ are particle positions, and $R$ is the arc distance. The natural length scale is $r_0=N_\text{p}^{-1/2}$, the magnetic length at half-filling. (For circular Fermi surfaces, $r_0$ coincides with $k_\text{F}^{-1}$ dictated by Luttinger's theorem). In parallel, we compute the static structure factor\cite{footnote_1}
\begin{gather}\label{eqn.static_fact}
 S_{l} = 1+ \dfrac{1}{N_\text{p}}\sum\nolimits_{a\neq b} \left\langle P_{l}(\vect r_a\cdot \vect r_b) \right\rangle~,
\end{gather}
where $P_{l}(x)$ are Legendre polynomials.

In a Fermi liquid, the structure factor is non-analytic near $2 l_{\text{F}}$, i.e., $S_{2 l_{\text{F}} +\delta l} - S_{2 l_{\text{F}}} \propto (\delta l)^{\alpha-3/2}$ with $\alpha_\text{FL}=3$. This singularity is reflected in a slow decay of $2k_\text{F}$ oscillations in real space, i.e., $g(R) \propto \cos(2 k_\text{F} R)/R^\alpha$ at long distances $\sqrt{N_\text{p}} \gg R/r_0 \gg 1$. In a spherical geometry, the oscillation amplitude increases as $R\rightarrow \pi$ due to constructive interference between different quasiparticle paths (see also App.~\ref{app.cfls}).

In strongly correlated metals with poorly defined quasiparticles, the exponent governing this decay can change, but power-law oscillations may persist even when quasiparticles become poorly defined.~\cite{Altshuler_low_energy_1994,Mross_controlled_expansion_2010} The presence of such oscillations is thus a useful numerical probe of emergent Fermi surfaces in quantum Hall systems and spin liquids.~\cite{Sheng_spin_bose_2009,Geraedts_half_filled_2016} When pairing gaps out the Fermi surface, the oscillations decay exponentially with a length scale $\xi_{2k_{\text{F}}} \propto \Delta_\text{pair}^{-1}$.

\begin{figure*}
\includegraphics[width=\textwidth,frame]{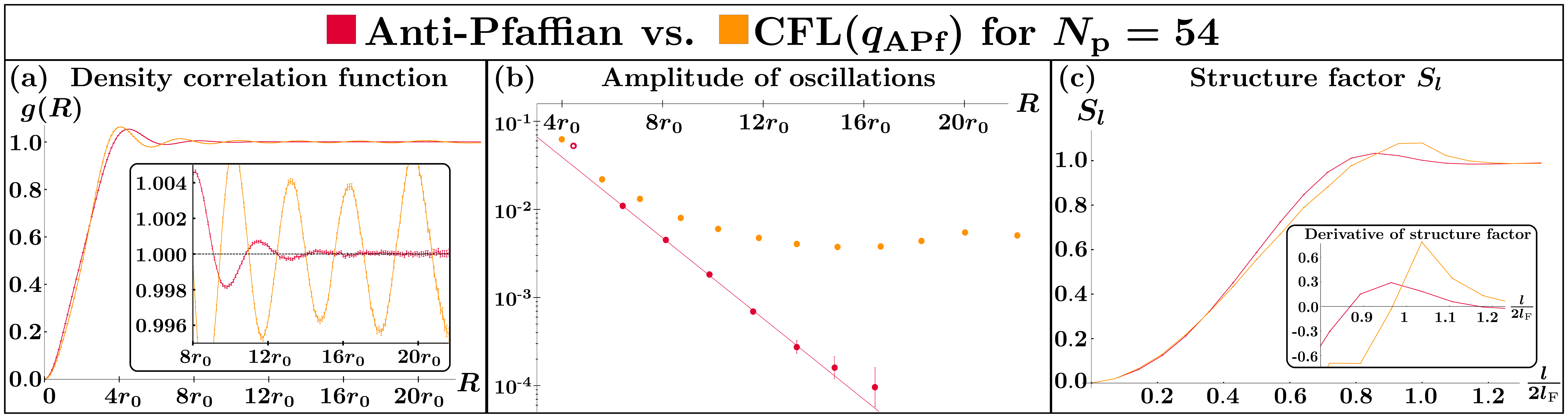}
\caption{\label{fig.APf}(a) The density-density correlation functions of the anti-Pfaffian and CFL trial states qualitatively match their analogs at $q_\text{MR}$, (cf.~Fig.~\ref{fig.cor_mr_cfl}). (b) The $2k_{\text{F}}$ oscillations decay exponentially with a length $\xi\approx 1.9 r_0$ for the anti-Pfaffian state but much slower for the CFL. As before, the $R\approx 4r_0$ peak is excluded from the fit. (c) The anti-Pfaffian structure factor is smooth, while that of the CFL exhibits a cusp at $2l_{\text{F}}$, similar to the one in Fig.~\ref{fig.cor_mr_cfl}(c).}

\end{figure*}

\subsection{Moore-Read Pfaffian}\label{sec.num.mr}
We begin our numerical analysis by revisiting the well-studied Moore-Read state. Specifically, we use a standard Monte Carlo algorithm to compute the density-density correlation function and the structure factor for the Moore-Read state and the CFL at the same monopole strength. Their behavior is well understood; we still reproduce them here as context for our subsequent analysis of the anti-Pfaffian and PH-Pfaffian states. To facilitate the comparison, we focus on particle numbers accessible for three states and that permit CFLs with filled shells, [cf.~Eq.~\eqref{filledshells}].

In Fig.~\ref{fig.cor_mr_cfl}(a), we show the density-density correlations for $N_\text{p}=56$. In the Moore-Read state, the initial oscillations decay rapidly, and $g(R)$ quickly approaches unity. A semilogarithmic plot of the oscillation amplitudes shows the exponential decay expected for a paired state [Fig.~\ref{fig.cor_mr_cfl}(b)]. We fit the decay lengths for the Moore-Read state with $N_\text{p}=20,30,42,56,72$ and extrapolate to the thermodynamic limit, where we find $\xi_{\text{MR}}=1.30(5)r_0$. This result is consistent with the value $\xi_\psi\approx 1.3r_0$, extracted in Ref.~\onlinecite{Moller_Neutral_2011,Bonderson_num_corrlength_2011} from the neutral fermion gap through exact diagonalization of Coulomb interactions. A somewhat longer length scale $\xi_\text{split}\approx 2.3r_0$ was obtained in Ref.~\onlinecite{Baraban_num_corrlength_2009} for the finite-size splitting between two putatively degenerate ground states.
 
In Fig.~\ref{fig.cor_mr_cfl}(c), we show the static structure factors. For the Moore-Read state, $S_l$ is smooth around $2l_{\text{F}}$. By contrast, it exhibits a cusp in the case of the CFL, consistent with the slow decay of real-space oscillations. Unfortunately, there is less than a decade between the short-distance peak near $R =4r_0$ and the onset of strong finite-size effects. Consequently, the exponent $\alpha$ cannot be obtained with confidence, and values in the range $2$\textendash$2.8$ are consistent with the data shown in Fig~\ref{fig.cor_mr_cfl}. Ref.~\onlinecite{Kamilla_static_1997} reported a best-fit value of $\alpha \approx 2.75$ for CFLs with $N_\text{p}=54$ particles at monopole strength $q_\text{aPf}=3/2$.

\subsection{Anti-Pfaffian}\label{sec.num.apf}

We now analyze the anti-Pfaffian trial state introduced in Sec.~\ref{sec.wf.apf}. As a first step, we compute its overlap with the explicit particle-hole conjugate of the Moore-Read state. The shifts of the two states imply that $\Psi_\text{anti-Pfaffian}(N_\text{p})=\hat \Theta \Psi_\text{Moore-Read}(N_\text{p}+2)$. Due to the exponential growth of the Hilbert space, we are only able to make an accurate comparison for up to ten particles. Tab.~\ref{tab.overlaps_apf} lists the overlaps between $\Psi_\text{anti-Pfaffian}$ and $\Psi^\text{CF}_\text{anti-Pfaffian}$ for any of the three projection methods described in Sec.~\ref{sec.LLL}. We find a remarkably large overlap above $99\%$ for two of the projections schemes. Unsurprisingly, the low-lying states in the entanglement spectrum agree even quantitatively (see App.~\ref{app.es.apfs}).

We now turn to a larger system of $N_\text{p}=54$, where a CFL with filled shells up to $l_\text{F}=13/2$ is also possible [cf.~Eq.~\eqref{filledshells}]. Here, only single and pairwise projections are applicable; we do not find significant differences between the two and quote numbers using the latter. We find the overlap between the anti-Pfaffian and CFL trial states to be $6.84(2)\%$. To compare this value to its analogue at the Moore-Read shift, one may look at states with the same Hilbert space size ($N_\text{p}=56$) or with the same $l_\text{F}$ ($N_\text{p}=42$). These overlaps are $1.3(1)\%$ and $4.8(5)\%$, respectively, comparable to those for the anti-Pfaffian.

Next, we compute the density-density correlations for anti-Pfaffian and CFL at $N_\text{p}=54$. Our results are shown in Fig.~\ref{fig.APf}; they closely mirror those of the Moore-Read state. For the anti-Pfaffian trial state, we find an exponential decay of oscillations and a smooth structure factor. We extract the correlation length for $N_\text{p}=28,40,54$ and extrapolate to the thermodynamic limit, where we find $\xi_{\text{aPf}}=1.38(14)r_0$, close to $\xi_{\text{MR}}=1.30(5)r_0$. The CFL wave function at the anti-Pfaffian shift behaves very similarly to the one at the Moore-Read shift, i.e., it exhibits a much slower decay of oscillations and a cusp at $2l_\text{F}$ in the structure factor.

Our findings provide strong evidence that the trial state $\Psi_\text{anti-Pfaffian}^\text{CF}$ can describe the anti-Pfaffian phase for large $N_\text{p}$. Its high overlap with the particle-hole conjugate of $\Psi_\text{MR}$ at moderate $N_\text{p}$ suggests that it may, moreover, be useful for addressing certain questions related to particle-hole symmetry. For example, the variational energies in the presence of three-body interactions (which can be attributed to Landau-level mixing) for Moore-Read and anti-Pfaffian states could be meaningfully compared.

\begin{table}
\caption{
\label{tab.overlaps_apf}
The anti-Pfaffian wave function Eq.~\eqref{eqn.APF} has a large overlap with the PH-conjugate of the Moore-Read state for $N_\text{p}=6,8,10$. The values for $\Psi^\text{parton}_\text{anti-Pfaffian}$ are from Ref.~\onlinecite{Balram_parton_2018}.}
\setlength{\extrarowheight}{4pt}
\begin{tabularx}{\columnwidth}{l |@{\hspace{5mm}} X X X }
 \hline \hline
Wave function & $N_\text{p} = 6$ & $N_\text{p} = 8$ & $N_\text{p} = 10$ 
 \\\hline
$\Psi^\text{CF}_\text{anti-Pfaffian}$ & $99.407(1)\%$ & $99.31(6)\%$ & $99.11(28)\%$
\\
$\Psi^\text{CF,single}_\text{anti-Pfaffian}$ & $99.260(18)\%$ & $99.36(8)\%$ & $99.12(16)\%$
\\
$\Psi^\text{CF,pair}_\text{anti-Pfaffian}$ & $97.925(6)\%$ & $98.20(9)\%$ & $95.99(14)\%$\\[2pt]
\hline
$\Psi^\text{parton}_\text{anti-Pfaffian}$ & $96.86\%$ & $95.23\%$ & $93.97\%$
\\[2pt]\hline\hline
\end{tabularx}
\end{table}

\subsection{PH-Pfaffian}\label{sec.num.ph}
\begin{figure*} 
\includegraphics[width=\textwidth,frame]{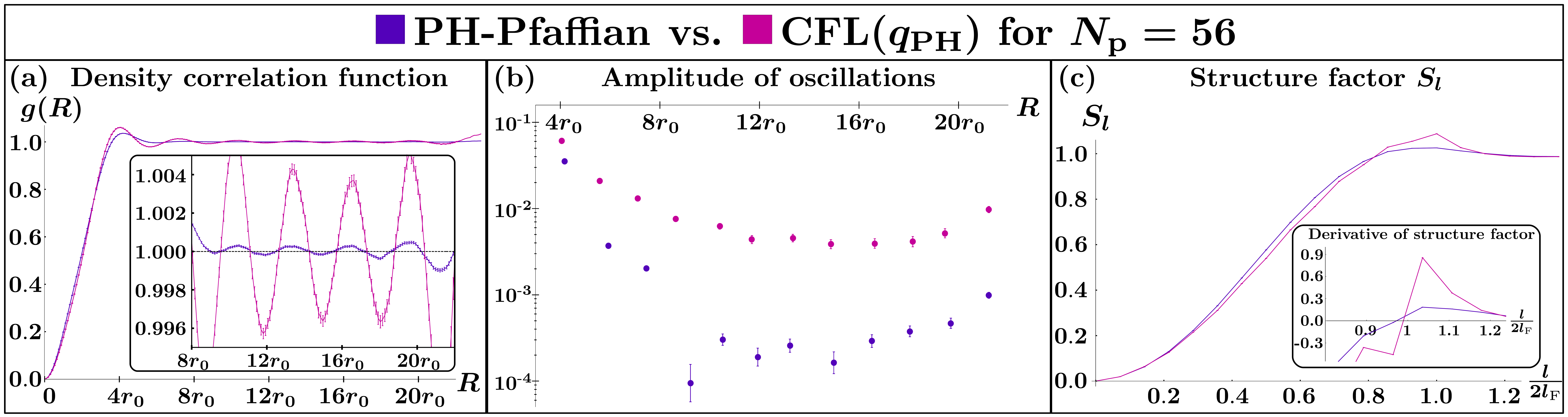}
\caption{ \label{fig.PH}
(a) Both the PH-Pfaffian trial state and the corresponding CFL exhibit slowly decaying $2k_\text{F}$ oscillations, with approximately the same wavelength. The overall amplitude of these oscillations is an order of magnitude smaller for PH-Pfaffian. (b) A semilogarithmic plot shows that the PH-Pfaffian oscillations are inconsistent with an exponential decay and instead behave similarly to those of the CFL. (c) The structure factor of the PH-Pfaffian state has a small cusp around $2l_{\text{F}}$, unlike those of Moore-Read and anti-Pfaffian.}
\end{figure*}

We begin our analysis of the PH-Pfaffian trial state by comparing the projection methods described in Sec.~\ref{sec.LLL} at small particle numbers. Up to $N_\text{p}=12$, all three projection methods are applicable and give similar results. The overlap between $\Psi_\text{PH}$ and its pairwise projected version is $99.2(5)\%$, somewhat larger than for single CF projection where we find $98.7(7)\%$. The overlap of all three states with the CFL is around $95\%$. For larger systems where only single and pairwise projection are feasible, we find no appreciable difference between the two, and that retaining CF Landau levels up to $2l_\text{F}$ is sufficient (see.~App.~\ref{app.PH_supp}). We therefore simply refer to $\Psi_\text{PH}^\text{pair}$ at this cutoff as `the PH-Pfaffian trial state.' For $N_\text{p}=56$, we find that its overlap with the CFL is still $~28.65(3)\%$, significantly larger than in the Moore-Read case where the analogous overlap is $ 1.3(1)\%$.

Next, we compute the density-density correlation functions for the PH-Pfaffian trial state and the CFL at the same shift (see Fig.~\ref{fig.PH}). We find that the $2 k_{\text{F}}$ oscillations of both states persist across the entire system [Fig.~\ref{fig.PH}(a)]. The non-universal \textit{overall} amplitude of oscillations is significantly smaller for the PH-Pfaffian trial state than for the CFL by an $R$-independent factor [Fig.~\ref{fig.PH}(b)]. (In a Fermi liquid, such an $R$-independent change of oscillation amplitudes may originate in a different quasiparticle weight.) The relatively weak oscillations are reflected in a rather faint cusp in $S_l$ [Fig.~\ref{fig.PH}(c)]. Notice, however, that both the PH-Pfaffian and the CFL trial states exhibit a peak in $\partial_l S_l$ at the same $l \approx 2l_\text{F}$---unlike the Moore-Read and anti-Pfaffian cases shown in Figs.~\ref{fig.cor_mr_cfl}(c) and \ref{fig.APf}(c). These observations strongly suggest that both wave functions lie in the same gapless CFL phase.

Finally, we perform a simple test of the relationship between the loss of a pairing gap and PH symmetry. We study the PH-Pfaffian trial state adapted to the filling factor $\nu=\frac{1}{4}$, where the question of PH symmetry does not arise. A possible trial wave function at this filling can be easily obtained by multiplying a $\nu=\frac{1}{2}$ wave function with a suitable Jastrow factor. However, such a wave function would presumably inherit many of the latter's properties, including any suppression of the gap. We instead choose $p=2$ in Eq.~\eqref{eqn.cforbital} and proceed with projection as described in Sec.~\ref{sec.LLL}. We find $2k_\text{F}$ oscillations that persist over the entire system for the largest system sizes that we studied ($N_\text{p}=30$); see App.~\ref{app.PH_supp} for details. This finding indicates that the obliteration of the gap due to projection is not necessarily related to PH symmetry, but this connection deserves a more systematic study.

\section{Summary and Conclusions}\label{sec.conc}

We have studied a class of paired-composite-fermions trial wave functions at filling factors $\nu = \frac{1}{2p}$. These wave functions can be efficiently projected into the LLL for any odd Cooper-pair angular momentum $\ell$. For $\ell=-1$, the Moore-Read wave function is reproduced exactly, which serves as a useful benchmark for subsequent approximations. The case $\ell=3$ describes a wave function in the anti-Pfaffian phase that, for moderate particle numbers, well approximates the particle-hole conjugate of the Moore-Read Pfaffian. This wave function may be useful in future comparative studies between these two phases, e.g., in the presence of Landau-level mixing.

The $\ell=1$ member of this family has been previously proposed to lie in the PH-Pfaffian phase.~\cite{Zucker_stabilization_2016} We have simulated this trial state at $\nu=\frac{1}{2}$ for relatively large system sizes of up to $N_\text{p}=56$ and found no evidence of a pairing gap on the composite-Fermi surface. To test for a possible relationship between the loss of a gap and PH symmetry, we further studied PH-Pfaffian states at $\nu=\frac{1}{4}$, and found similar behavior to the half-filled case.

The variational freedom afforded by the wave function of Eq.~\eqref{eqn.MS} may well permit a fully gapped PH-Pfaffian phase in the lowest Landau level. We note that a pure power-law dependence $\eta_l \sim l^{-1 +\gamma}$ in Eq.~\eqref{eqn.Phi} corresponds to a Cooper-pair wave function $g(R)\sim R^{-1-\gamma}$ for $\gamma \in [0,1 ]$. Insisting on weak-pairing behavior of the unprojected wave function fixes $\gamma =0$ and a scale-dependent choice of the parameters $\eta_l$ is required to access different states. Here, $l_\text{F}$ constitutes the only available scale, and larger values $\eta_{l}$ for $l > l_\text{F}$ promote pairing correlations. For the \textit{projected} wave function, we have found that only parameters with $l \leq 2l_\text{F}$ play a significant role. There are consequently a relatively small number of variational parameters, and it would be interesting to explore whether they permit access to the fully gapped PH-Pfaffian phase.

Finally, we have observed that different means of projecting the same CF ansatz can result in LLL wave functions that describe altogether different phases of matter. Depending on the specific trial state and the purpose of the study, either method may be preferable. The pairwise projection method introduced here provides an alternative to the widely used single CF projection and is, likewise, suitable for large-scale Monte Carlo simulations.

\begin{acknowledgments}
It is a pleasure to thank Jason Alicea, Tobias~Holder, Ryan Mishmash, and Olexei~I.~Motrunich for illuminating discussions. This work was supported by the Israel Science Foundation (ISF) and the Minerva Foundation with funding from the Federal German Ministry for Education and Research.
\end{acknowledgments}

\appendix

\section{Mean-field-superconductor wave function}
\label{app.SC_WF}
The wave functions of spinless mean-field superconductors can be constructed as prescribed in Ref.~\onlinecite{Read_paired_2000}. Consider a model of spinless fermions in two dimensions with a mean-field pairing term, i.e.,
\begin{flalign}\label{eq: H sup}
H_\text{BCS}=\sum\nolimits_{\vect k}\left[ \xi_{\vect k}c^{\dag}_{\vect{k}}c_{\vect{k} } +\frac{1}{2}(\Delta_{\vect{k}}^*c_{-\vect{k}}c_{\vect{k}}+\text{H.c.}) \right]~.
\end{flalign} 
The Hamiltonian $H_\text{BCS}$ is diagonalized by the Bogoliubov transformation $a_{\vect{k}}=u_{\vect{k}}c_{\vect{k}}-v_{\vect{k}}c_{-{\vect{k}}}^{\dag}$, i.e.,
\begin{equation}\label{eq: H diag}
 H_\text{BCS}=\sum\nolimits_{{\vect k}}E_{{\vect k}}a^{\dag}_{\vect{k}}a_{\vect{k}}~,\qquad E_{\vect k}=\sqrt{\xi_{\vect k}^2+|\Delta_{\vect{k}}|^2}~.
\end{equation} 
The functions $u_{\vect{k}}$ and $v_{\vect{k}}$ (not to be confused with the coordinates $u_a,v_a$) satisfy $|u_{\vect{k}}|^2+|v_{\vect{k}}|^2=1$ and
 \begin{equation}\label{eq: def g}
 g_{\vect{k}} \equiv \dfrac{v_{\vect{k}}}{u_{\vect{k}}}=\dfrac{\xi_{\vect k}-E_{\vect k}}{\Delta^*_{\vect{k}}}~.
 \end{equation}
 The ground state $|\Omega\rangle$ of $H_\text{BCS}$ satisfies $a_{\vect{k}}|\Omega\rangle=0$ for all $\vect{k}$; in terms of the original fermions $c^{\dag}_{\vect{k}}$ it is
 \begin{equation}\label{eq: exp}
 |\Omega\rangle
 \propto \exp\left[\dfrac{1}{2}\sum\nolimits_{\vect k}g_{\vect{k}}c^{\dag}_{\vect{k}}c^{\dag}_{-\vect{k}}\right] |0\rangle~.
\end{equation}
The corresponding real-space wave function for an even number of fermions is given by
\begin{flalign}\label{eq: Pf}
\Psi(\vect{r}_{1},\ldots,\vect{r}_{N_\text{p}})=\text{Pf}[g(\vect{r}_a-\vect{r}_b))]~,
\end{flalign}
where $g(\vect{r})$ is the Fourier transform of $g_{\vect{k}}$.

For spinless fermions, $\Delta_{\vect{k}}$ must be an odd function of ${\vect{k}}$ and, in particular, vanishes at $\vect k=0$. Thus, we consider pairing with odd angular momentum $\ell$, i.e., $\Delta_{\vect{k}} =\Delta|\vect{k}|e^{i\ell \vartheta_{\vect{k}}}$, where $\vartheta_{\vect{k}}$ is the angle between $\vect{k}$ and the $x$-axis. Using Eq.~\eqref{eq: def g}, one finds $\lim \nolimits_{|\vect k| \rightarrow 0}g_{\vect{k}}= -\frac{2\mu}{\Delta_{\vect{k}}^*}$ for positive chemical potential. 
The choice of $\Delta_{\vect{k}}$ above results in $g_{\vect{k}}\sim-\frac{2\mu}{\Delta}\frac{1}{|\vect k|}e^{i\ell \vartheta_{\vect{k}}}$ and the Fourier transform
\begin{align}
g(\vect{r}) =
 -i^{\ell} e^{i \ell \vartheta_{\vect{r}}}\int dk k|g_{\vect{k}}| J_{\ell}(kr)~,
\end{align}
 where $J_\ell$ is the Bessel function of the first kind, which satisfies the normalization $\int_{0}^\infty dx J_\ell(x)=1$. The momentum integral converges rapidly enough that is suffices to insert the small-$\vect{k}$ limit of $g_{\vect{k}}$. We thus find
\begin{align}\label{eq: g_l}
g(\vect{r}) \approx \dfrac{2\mu}{\Delta}\frac{-i^{\ell}}{|\vect r|}e^{i\ell\vartheta_{\vect{r}}}~,
\end{align}
precisely the pair wave function in Eq.~\eqref{eqn.addition}.

\section{Expansion of Cooper pair wavefunction in monopole harmonics}
\label{app.addition}
To derive Eq.~\eqref{eqn.addition}, first start with the addition theorem for monopole harmonics, Refs.~\onlinecite{Wu_dirac_1976,Wu_properties_1977}, which can be expressed as
\small
\begin{align}\label{eq.b1}
 \sum_{m=-l}^{l}Y^{q}_{l,m}(\vect r_a) {\overline{Y}}^{q}_{l,m}(\vect r_b)={ \sqrt{\frac{2l+1}{4\pi}}}Y^{q}_{l,q}(\theta,\phi=0)F_{q}(\vect r_a,\vect r_b)~.
\end{align}\normalsize
Here the angle $\theta$ is given by $\vect r_a \cdot \vect r_b = \cos \theta = 1 -2 |\omega_{ab}|^2$ and
$F_{q}(r_a,r_b)$ is a phase factor that does not depend on the Landau-level index $l-q$. To determine this factor, it is sufficient to focus on the LLL, $l=q$, where considerable simplifications occur. Specifically, for positive $q$ the monopole harmonics satisfy
\small
\begin{align}
 Y_{q,m}^q(\vect r_a){\overline Y}_{q,m}^{q}(\vect r_b)=
{\frac{2q+1}{4\pi}} \begin{pmatrix}2q\\q-m\end{pmatrix} (u_av_b)^{q-m}(u_bv_a)^{q+m}~,
\end{align}
\normalsize
and thus
\small
\begin{align}
&\sum_{m=-q}^{q} Y_{q,m}^q(\vect r_a){\overline Y}_{q,m}^{q}(\vect r_b)= \frac{2q+1}{4\pi}(u_av_b-u_bv_a)^{2q}~. 
\end{align}
\normalsize
Inserting this phase factor into Eq.~\eqref{eq.b1} yields
\begin{align}
\sum_{m=-l}^{l}Y^{q}_{l,m}(\vect r_a) \overline{Y}^{q}_{l,m} (\vect r_b)= \frac{2l+1}{4\pi}\omega_{ab}^{2q} P_{l-q}^{(2q,0)}(\cos \theta),
\label{eq.b4}
\end{align}
where $P_{n}^{(2q,0)}$ are Jacobi polynomials, which satisfy
\begin{flalign}
\frac{1}{|\omega_{ab}|^{2q+1}}=2\sum_{n=0}^{\infty}P_{n}^{(2q,0)}(\cos \theta)~.
\label{eqn.generating_functional}
 \end{flalign} 
 (This relation follows directly from the generation functional). Multiplying Eq.~\eqref{eqn.generating_functional} with $\omega_{ab}^{2q}$ and using Eq.~\eqref{eq.b4}, we arrive at the relation quoted in the main text, i.e.,
\begin{subequations}
\begin{align}
 \dfrac{\omega_{ab}^{q-1/2}}{\bar{\omega}_{ab}^{q+1/2}}
 &=2\omega_{ab}^{2q}\sum_{n=0}^{\infty}P_{n}^{(2q,0)}(1-2|\omega_{ab}|^2) 
 \label{eqn.A_expanssion1}\\ 
 &=\sum^{\infty}_{l=|q|}\dfrac{8\pi}{2l+1}\sum_{m=-l}^{l}Y^{q}_{l,m}(\vect r_{a}) \overline{Y}^{q}_{l,m} (\vect r_{b}).
 \label{eqn.A_expanssion2}
\end{align}
\end{subequations}
Using complex conjugation and symmetries of the monopole harmonics, one finds that Eq.~\eqref{eqn.A_expanssion2} also holds for negative monopole strength $q$.

\section{Details on pairwise projection}
\label{app.Pairwise_projection}

Any spherically symmetric composite-fermion-pair wave function that may appear in the argument of the Pfaffian in Eq.~\eqref{eqn.CF_pair_form} can be expressed using Eq.~\eqref{eq.b4} as
\begin{align}
\label{eqn.c1}
g^\text{CF}_{ab} = \sum_{n} C_n\bar{\omega}_{ab}^{n} \omega^{n}_{ab}\Phi^{Q}~,
\end{align}
where $\Phi^Q$ is a LLL wave function at monopole strength $Q>0$. This function can be projected by replacing $\bar \omega_{ab}$ with the differential operator $\hat d_{ab}\equiv \partial_{u_a}\partial_{v_b}-\partial_{u_b}\partial_{v_a}$, i.e.,
\begin{align}
\PLL\left\{ \bar{\omega}_{ab}^n \omega_{ab}^{n}\Phi^Q \right\}=\left[ \dfrac{(2Q+1)!}{(2Q+n+1)!} \right]^2 \hat{d}_{ab}^n\omega_{ab}^{n}\Phi^{Q}.
\label{eq.c2}
\end{align} 
To commute all $d_{ab}$ to the right of all $\omega_{ab}$, we introduce the operator $\hat{t}_{ab}\equiv u_a\partial_{u_a}+v_a\partial_{v_a}+u_b\partial_{u_b}+v_b\partial_{v_b}$, which satisfies $\hat t\Phi^Q =4 Q \Phi^Q$. Moreover, $\hat d$, $\hat{t}$, and $\omega$ form a closed algebra
\begin{align}
[\hat d,\omega]=\hat t+2~,\quad [\hat t,\omega]=2\omega~,\quad [\hat t,\hat d]=-2\hat d~.
\end{align}
[]This is a representation of $SU(2)$ with the identification $L_+=\omega$, $L_-=-d$, and $L_z=\frac{\hat{t}+2}{2}$]. After a straightforward calculation, we find
\begin{align}
\hat{d}^s\omega^n =\sum_{k=0}^{\text{min}\{n,s\}}\dfrac{n!s!}{k!}\dfrac{\omega^{n-k}}{(n-k)!}\dfrac{\hat d^{s-k}}{(s-k)!}\hat{f}^{n-s}_{k},\label{eq.c4}
\end{align}
where $\hat{f}^{\delta}_{k}\equiv \prod_{i=1}^{k}(\hat{t}+i+\delta+1)$. For CFs at filling factor $\nu=1/2p$, the function $\Phi^Q$ is given by
\begin{align}
\Phi^Q= \omega_{ab}^{2p+2q} {\cal J}_{ab}^p\equiv \omega_{ab}^{2p+2q}\left(\prod\nolimits_{i\neq a,b} \omega_{ai}\omega_{bi} \right)^p~.
\end{align}

\begin{figure}[b]
\includegraphics[width=.92\linewidth,frame]{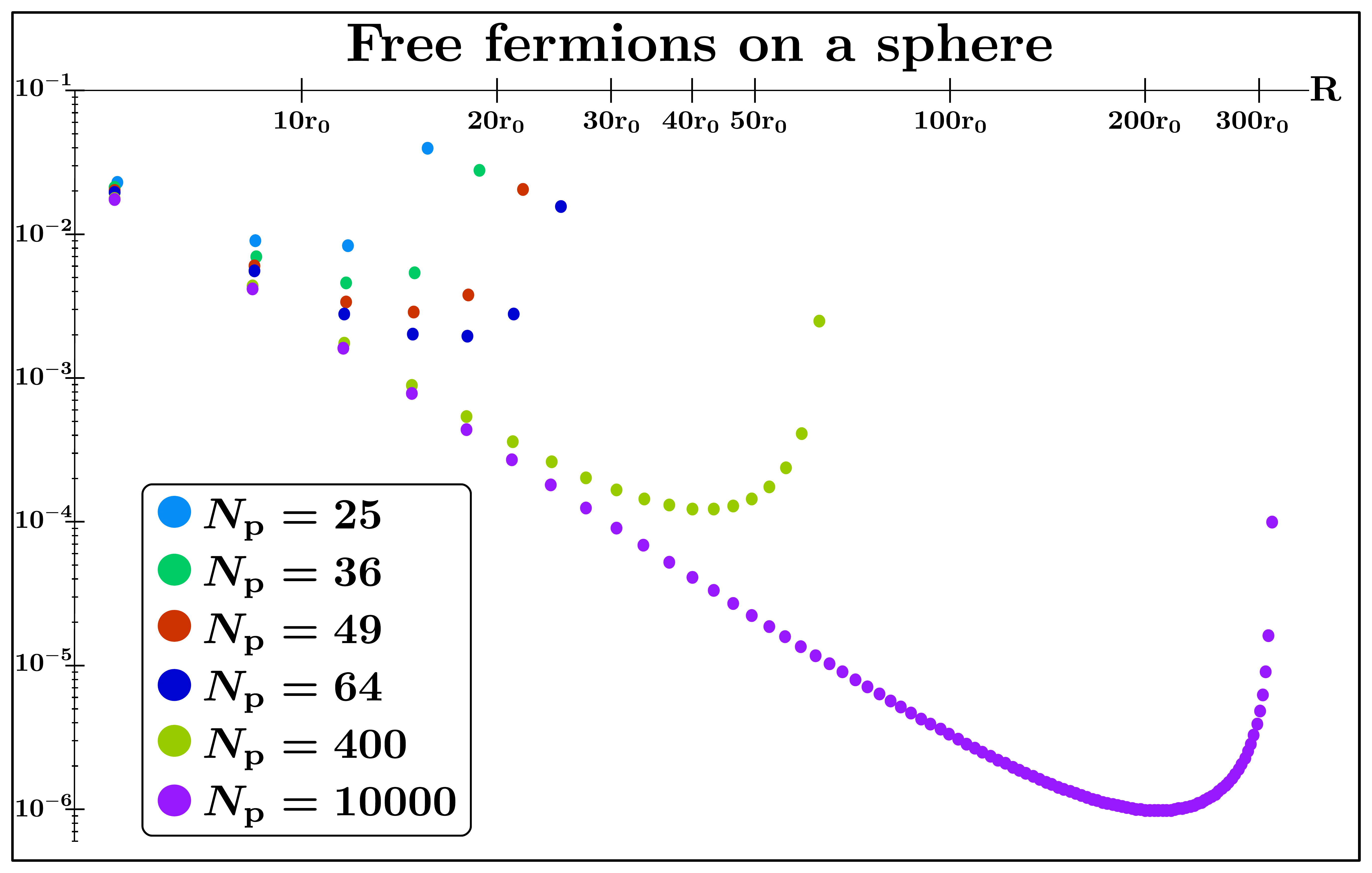}
\caption{\label{fig.FFS} The amplitude of $2k_\text{F}$ oscillations exhibited by free fermions on a sphere deviates significantly from the $N_\text{p}\rightarrow \infty$ limit (solid line) up to relatively large particle numbers.}
\end{figure}

To perform projection, one still needs to compute the derivatives $d_{ab}^n\mathcal{J}_{ab}^p$. In the case $p=1$ the expressions simplify considerably. We observe that
\begin{align}
\hat{d}_{ab}(\omega_{ai}\omega_{bj})=\omega_{ij}=\Omega_{ij}\omega_{ai}\omega_{bj}~,\quad\Omega_{ij}\equiv\dfrac{\omega_{ij}}{\omega_{ai}\omega_{bj}}~,
\end{align}
and compute
\begin{align}
\hat d_{ab}\mathcal{J}_{ab}&=\hat d_{ab}\left( \prod\nolimits^{\prime}_{i}\omega_{ai}\omega_{bi} \right)
=
\mathcal{J}_{ab}\sum\nolimits_{i,j}^{\prime}\Omega_{i,j}\nonumber\\&=
\sum\nolimits_{i,j}\omega_{ij}\left( \prod^{\prime}\nolimits_{k\neq i,l\neq j }\omega_{ak}\omega_{bl} \right)~.
\label{eqn.B_d}
\end{align}
Here, the prime specifies that indices run over all particles other than $a$ and $b$. Notice that the product in the last line of Eq.~\eqref{eqn.B_d} has a similar structure as $\mathcal{J}_{ab}$. Consequently, acting repeatedly with $\hat d_{ab}$ results in
\begin{align}\label{eq.c8}
\hat{d}_{ab}^n\mathcal{J}_{ab}=\mathcal{J}_{ab}\sum_{i_1\neq\ldots i_n}\sum_{j_1\neq\ldots j_n}\prod_{\alpha=1}^{n}\Omega_{i_\alpha j_{\alpha}}~.
\end{align}
Next, we introduce $h^{ab}_i=\omega_{ai}/\omega_{bi}$ to rewrite
\begin{align}
\omega_{ab}\Omega_{ij}=1-h^{ab}_i h^{ba}_j~.\label{eq.c9}
\end{align}
As a final step, we multiply Eq.~\eqref{eq.c8} by $\omega_{ab}^n$ and use Eq.~\eqref{eq.c9} to find
\begin{flalign}\label{eqn.dJ}
\omega^{n}_{ab}\hat{d}_{ab}^n\mathcal{J}_{ab}=\sum_{k=0}^{n}(-1)^k\begin{pmatrix}
n \\
k
\end{pmatrix}\left( \dfrac{k!(N-k)!}{(N-n)!} \right)^2\mathcal{J}_{ab}e^{ab}_{k}e^{ba}_{k}.
\end{flalign}
Here $N=N_\text{p}-2$ and $e^{ab}_k$ are the elementary symmetric polynomial in $N$ variables $h_i$ for $i\neq a,b$, i.e.,
\begin{flalign}\label{eqn.sym_ploy}
e^{ab}_{k}(\{h^{ab}_j\})=\sum_{i_1<\ldots< i_k}\prod_{\alpha=1}^{k}h^{ab}_{i_\alpha}~.
\end{flalign}
A straightforward evaluation of these sums would require on the order of $N_\text{p}^n$ operations, but can fortunately be avoided. Refs.~\onlinecite{Davenport_projection_2012,Mukherjee_incompressible_2015,Fulsebakke_projection_2016} developed an efficient algorithm for evaluating $e^{ab}_n$, which can be readily adapted to the present case.\footnote{One can always find a conformal transformation that maps $\vect{r}_a$ and $\vect{r}_b$ to opposite poles and thus $h_i \rightarrow u_i/v_i$, while the cross-ratio $\omega_{ab}\Omega_{ij}$ is conformally invariant.} In our simulations we further use the library of Ref.~\onlinecite{pfapack_Wimmer_2012} for efficiently evaluating Pfaffians.

For the wave functions introduced in the main text, the steps described by Eqs.\eqref{eqn.c1}\textendash\eqref{eq.c4} result in
\small
\begin{align}\PLL\left[ \dfrac{\omega_{ab}^{q-1/2}}{\bar{\omega}_{ab}^{q+1/2}}\omega_{ab}^{2p}\mathcal{J}_{ab}^{p} \right]_{N_c}=\sum_{n=0}^{N_\text{c}}A^{N_\text{c}}_{n}\omega_{ab}^{n+2p+2q}d_{ab}^n\mathcal{J}_{ab}^p
,\label{eqn.sum_pair_WF}
\end{align}
\normalsize
where $N_\text{c}=l_\text{c} - |q|$ is the cutoff [cf.~Eq.~\eqref{eqn.MR_exp}], and the coefficients $A^{N_\text{c}}_{n}$ for $q>0$ are given by
\small
\begin{align}
A^{N_\text{c}}_{n}=
\sum_{s=n}^{N_\text{c}}\sum_{k=n}^{s}\dfrac{2(-1)^k}{n!(s-k)!}\dfrac{[2p(N_\text{p}-1)+2q+1+k-n]!}{[2p(N_\text{p}-1)+2q+1]!(k-n)!}\nonumber
\\\times
\dfrac{(s+k+2q)!}{(k+2q)!}
\dfrac{[k+2(p+q)]!}{[n+2(p+q)]!}
\left[ \dfrac{(p N_\text{p}+2q+1)!}{(p N_\text{p}+k+2q+1)!} \right]^2.
\end{align}
\normalsize

\begin{figure}[t]
\includegraphics[width=0.9\linewidth,frame]{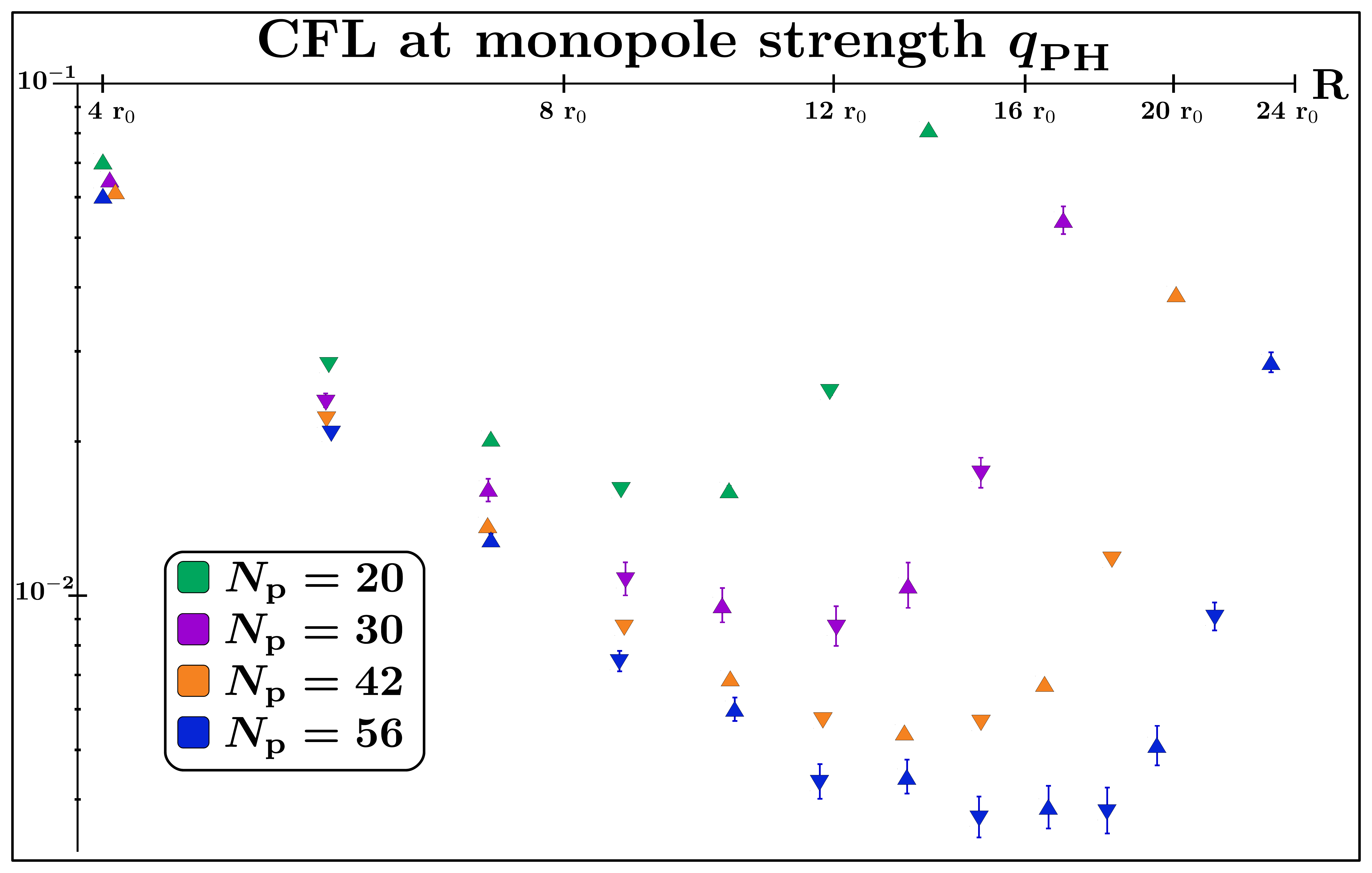}
\caption{\label{fig.CFL} The $2k_\text{F}$ oscillations of the CFL at $q_\text{P}$ qualitatively match those of free-fermions. Notice that the odd peaks are systematically higher than the even ones, since we show both maxima (up-pointing triangles) and minima (down-pointing triangles), and the base-line is non-constant.}
\end{figure}

\begin{figure}[b]
\includegraphics[width=0.9\linewidth,frame]{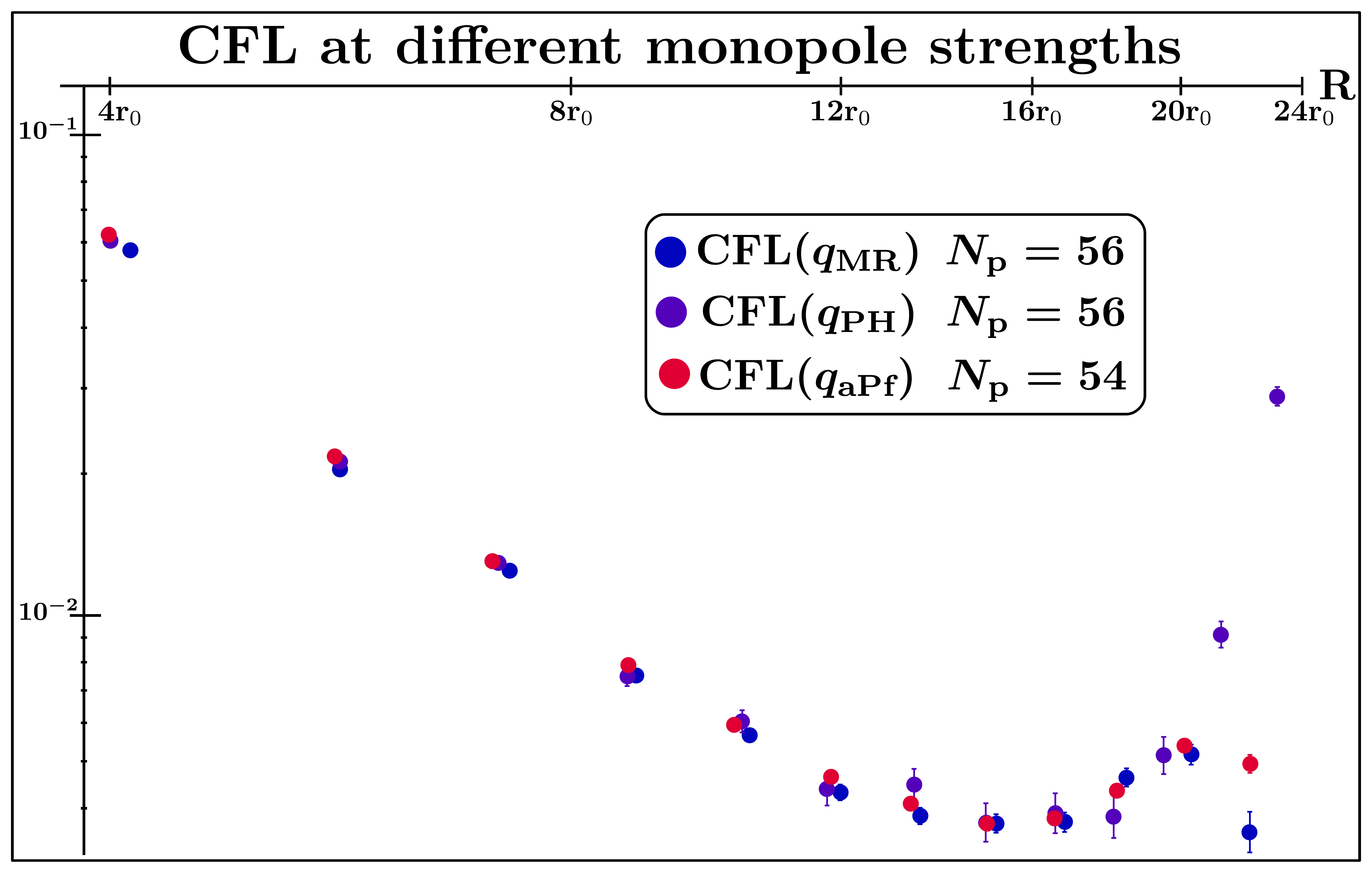}
\caption{\label{fig.CFL_shift} CFLs at different monopole strengths and comparable particle numbers exhibit similar $2k_\text{F}$ oscillation apart from the final peak.}
\end{figure}

\section{Fermi liquids and composite Fermi liquids on a sphere}\label{app.cfls}

To help interpret our numerical results for CFLs, it is instructive to recall the free-fermion behavior in the same geometry. At monopole strength $q=0$, the free-fermion structure factor is given by
\begin{align}
 S_{l}=1-\sum_{l_1,l_2=0}^{l_\text{F}}\dfrac{(2l_1+1)(2l_2+1)}{4\pi(l_\text{F}+1)^2} \begin{pmatrix}
 l & l_1 & l_2\\ 0&0&0
 \end{pmatrix}^2,
\end{align}
where the last term on the right-hand side is the Wigner 3$j$ symbol. The amplitudes of the corresponding $2k_\text{F}$ oscillations~\cite{footnote_1} are shown in Fig.~\ref{fig.FFS}. As expected for a gapless state, there are strong finite-size effects. Even for large system sizes inaccessible by Monte Carlo methods, there are significant deviations from the thermodynamic behavior.  In particular, the oscillation amplitudes \textit{increase} as $R \rightarrow \pi$, i.e., between antipodal points.  A numerically accurate determination of the exponent $\alpha$ is thus challenging.

For the CFLs, we obtained data for up to $N_\text{p}=56$. We find good qualitative agreement with the free fermion behavior (Fig.~\ref{fig.CFL}). In particular, the decay of oscillations changes substantially between $N_\text{p}=20$ and $N_\text{p}=56$. Any decay exponent extracted from small to moderate system sizes should thus be viewed as a lower bound on its actual value. Still, our data suggest that $\alpha$ takes a somewhat smaller value than for free fermions in the thermodynamic limit.

We find only a weak dependence of the $2k_\text{F}$ oscillation on the monopole strength (Fig.~\ref{fig.CFL_shift}). The data for $q_\text{MR}$, $q_\text{aPf}$, and $q_\text{PH}$ deviate only very close to $R\approx \pi$. The latter exhibit an upturn that corresponds to constructive interference, similar to the case of free fermions at zero monopole strength. By contrast, there appears to be destructive interference for 
$q_\text{MR}$ and $q_\text{aPf}$ (notice, however, that the final dip is preceded by an increase for $R \gtrsim \pi/2)$.

\begin{table}[t]
\caption{
\label{tab.convergence_PH}
 At $N_\text{p}=56$, the PH-Pfaffian trial wave function with pairwise projection is converged to more than $99\%$ at the cutoff $N_\text{c}= l_\text{c} - |q|\geq 14$; for single CF projection, the convergence is reached for $N_\text{c} \geq 12$. }
\setlength{\extrarowheight}{3pt}
 \begin{tabularx}{\linewidth}{r | >{\raggedright\arraybackslash}X >{\raggedright\arraybackslash}X >{\raggedright\arraybackslash}X >{\raggedright\arraybackslash}X}
 \hline
\hline
 &\ \ $N_\text{c} =12$ & $N_\text{c} = 14$ &$N_\text{c} = 15$ \\
\hline
 $|\langle\Psi^{\text{pair}}_{\text{PH},N_c=20}|\Psi^{\text{pair}}_{\text{PH},N_\text{c}}\rangle|$ \ 
 &\ \ $ 88.7(2)\%$ & $99.42(2)\%$ & $99.90(1)\%$ 
 \\
 $|\langle\Psi^{\text{single}}_{\text{PH},N_c=20}|\Psi^{\text{single}}_{\text{PH},N_\text{c}}\rangle|$ \
 &\ \ $99.79(1)\%$ & $99.980(1)\%$ & $99.9994(1)\%$ \\[1.5pt]
\hline\hline
\end{tabularx}

\end{table} 
 \begin{figure}[b]
\includegraphics[width=0.9\linewidth,frame]{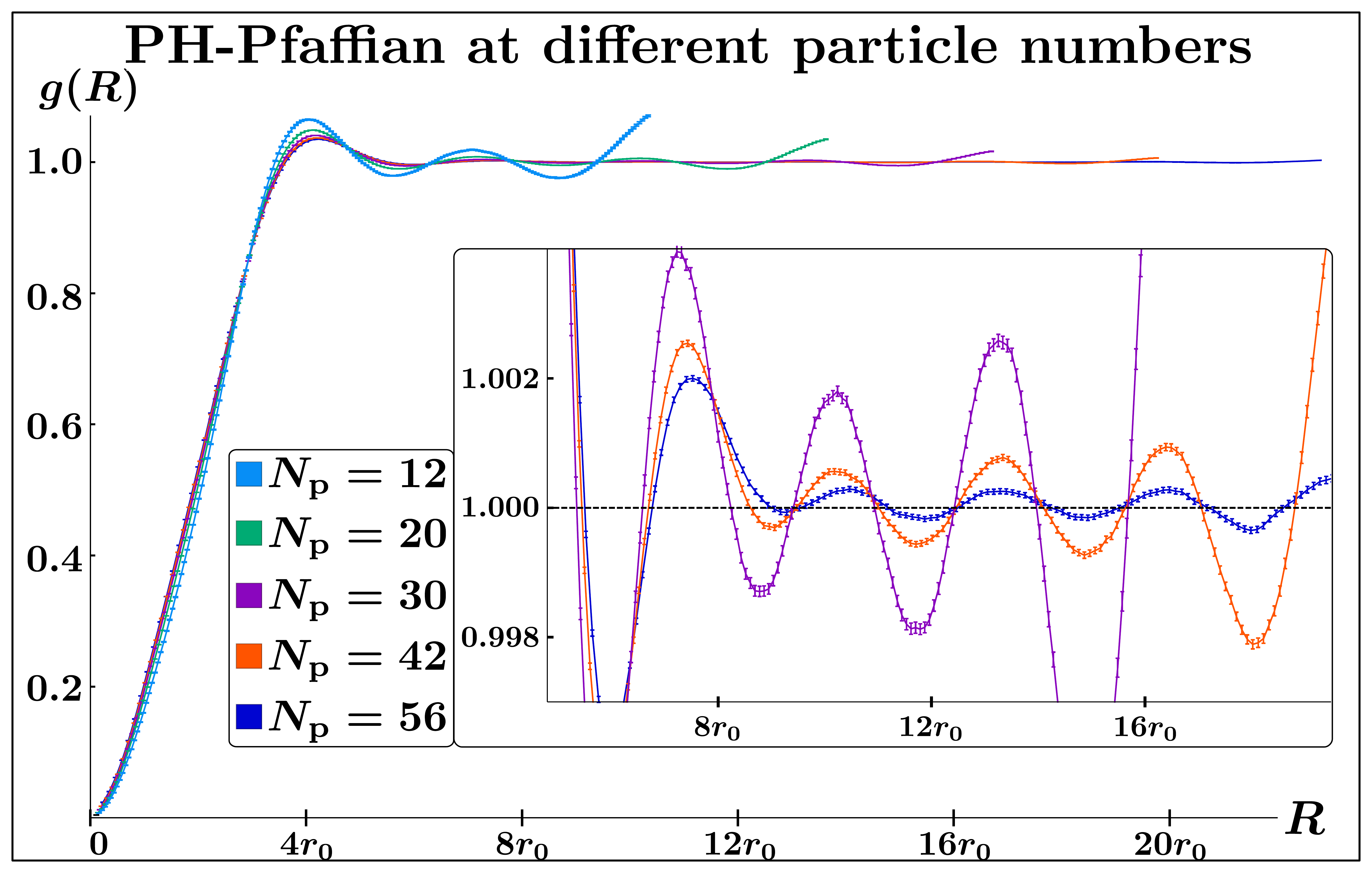}
\caption{ \label{fig.ph_56-30}
The density-density correlation function of the PH-Pfaffian trial state shows a global suppression of the $2k_\text{F}$ amplitudes with increasing particle number, but almost no change in the $R$ dependence.}
\end{figure}

\begin{figure}[t]
\includegraphics[width=.9\linewidth,frame]{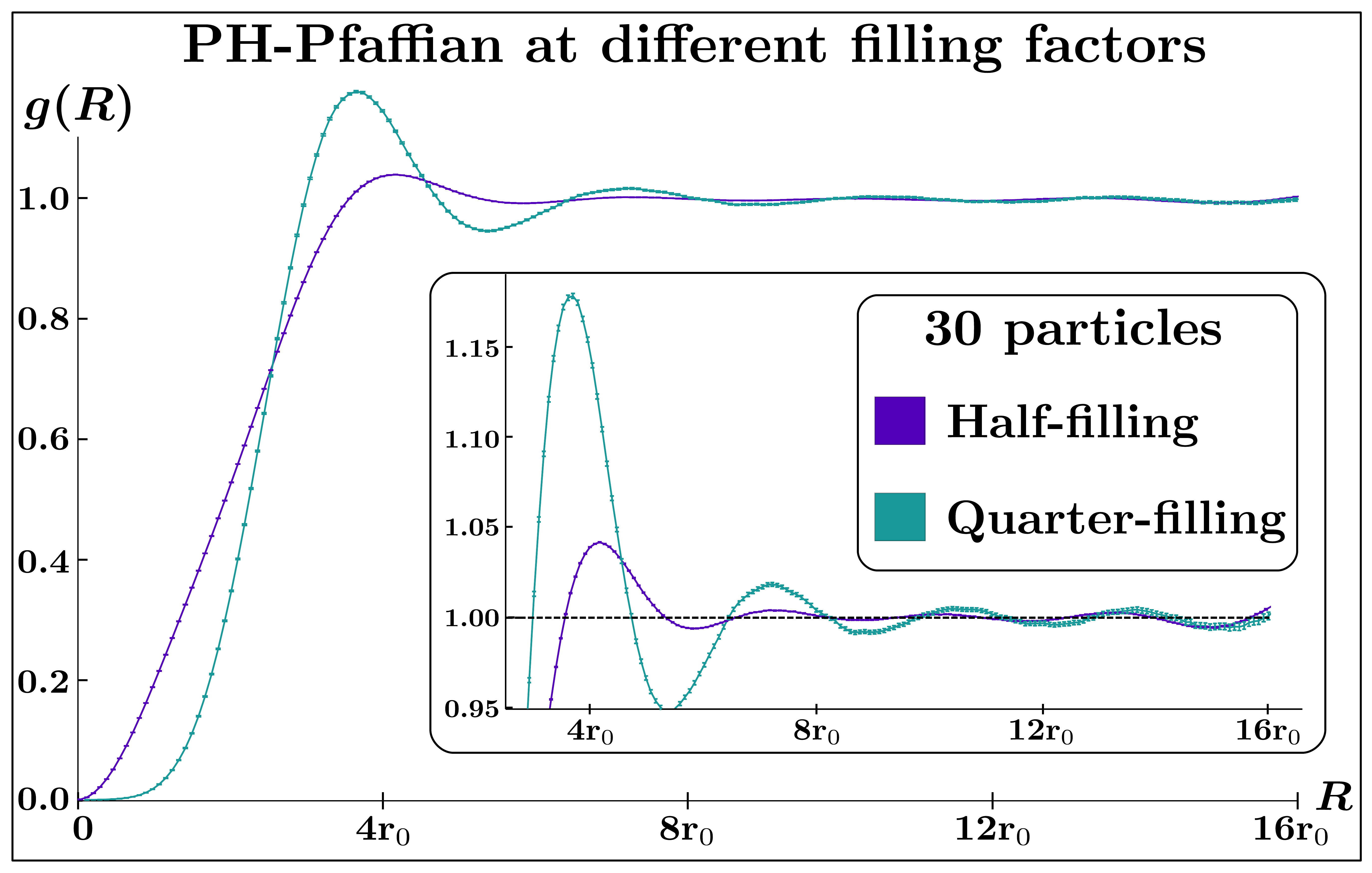}
\caption{\label{fig.ph_nu_1_4}The density-density correlation functions of the PH-Pfaffian with $N_\text{p}=30$ at filling factors $\nu=\frac{1}{4}$ and $\nu=\frac{1}{2}$ are qualitatively similar. At short distances $g(R) \propto R^{2/\nu-2}$, and $2k_\text{F}$ oscillations persist for all $R$.}
\end{figure}

\section{PH-Pfaffian supplementary data}\label{app.PH_supp} 

We show the convergence of the single and pairwise projection methods in Tab.~\ref{tab.convergence_PH}. Specifically, we take the pairwise projected state with $N_\text{c}= l_\text{c} - |q|=20$ as a reference for $N_\text{p}=56$ and determine the overlap for any smaller cutoff. We find that the convergence is exponentially fast, and an overlap above $99\%$ is reached for $N_c \geq 14$. In general, the cutoff $l_\text{c}\approx 2l_\text{F}$ provides a good approximation at any particle numbers.

To further characterize the PH-Pfaffian trial state, we show the density-density correlation function for different particle numbers in Fig.~\ref{fig.ph_56-30}. The overall $R$-independent amplitude of oscillations decreases with particle number, but there is no significant difference in the $R$ dependence. For all $N_\text{p}$, the oscillation amplitudes exhibit an increase for $R\gtrsim\pi/2$, similar to free fermions and CFLs [cf.~App.~\ref{app.cfls}]. Consequently, we may place the bound $\xi_\text{PH}>12 r_0$ on the correlation length of the PH-Pfaffian trial state. This value is an order of magnitude larger than those found for the Moore-Read and anti-Pfaffian wave functions.

Finally, we compare the PH-Pfaffian trial states at filling factors $\nu=\frac{1}{2}$ and $\nu=\frac{1}{4}$ for $N_\text{p}=30$ in Fig.~\ref{fig.ph_nu_1_4}. In the quarter-filled case, there is a somewhat stronger $R$ dependence of the oscillation amplitude, but at both fillings, the oscillations persist over the entire sphere. It may be worth exploring whether the $R$ dependence is indicative of an exponential decay at larger $N_\text{p}$, but the correlation length would still be substantially longer than in the case of Moore-Read or anti-Pfaffian.

\begin{figure}[b]
\centering
 \fbox{\includegraphics[width=0.95\linewidth]{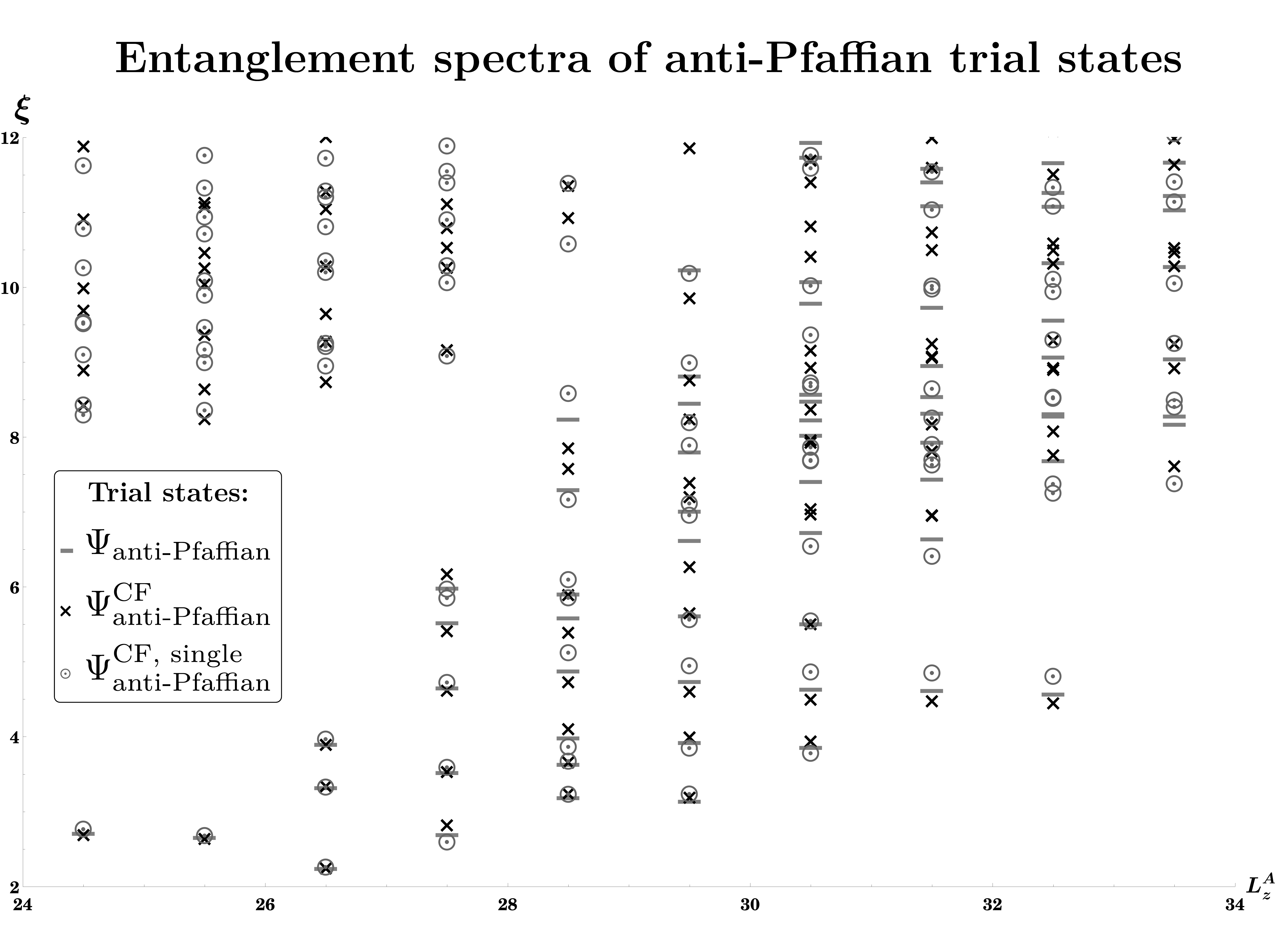} }
 \caption{ \label{fig.ES_APf}  The low-lying states in the orbital entanglement spectra of the trial states $\Psi^\text{CF}_\text{anti-Pfaffian}$, and $\Psi^\text{CF, single}_\text{anti-Pfaffian}$ exhibit perfect qualitative agreement with those obtained through PH conjugation of the Moore-Read wave function. }
\end{figure}

\section{Anti-Pfaffian entanglement spectra}\label{app.es.apfs}
The trial states $\Psi^\text{CF}_\text{anti-Pfaffian}$ and $\Psi_\text{anti-Pfaffian}$ exhibit overlaps above $99\%$ for $N_\text{p}\leq 10$ with the PH-conjugate Moore-Read wave function (see Tab.~\ref{tab.overlaps_apf}). Thus, it is unsurprising that their entanglement spectra~\cite{Li_entanglement_spectrum_2008} also match well, but we still provide them for completeness. Specifically, we perform an orbital decomposition where subsystem $A$ contains five particles with positive angular momentum $L_z$ and $B$ the other five with negative $L_z$. In Fig.~\ref{fig.ES_APf}, we plot the corresponding entanglement energies as a function of the total angular momentum in subsystem $A$. When using pairwise projection, the overlaps with $\Psi_\text{anti-Pfaffian}$ are somewhat smaller, but the entanglement spectrum still matches qualitatively; the degeneracies of the low-lying states are identical.

\bibliography{phpfbib}

\begin{thebibliography}{82}%
\makeatletter
\providecommand \@ifxundefined [1]{%
 \@ifx{#1\undefined}
}%
\providecommand \@ifnum [1]{%
 \ifnum #1\expandafter \@firstoftwo
 \else \expandafter \@secondoftwo
 \fi
}%
\providecommand \@ifx [1]{%
 \ifx #1\expandafter \@firstoftwo
 \else \expandafter \@secondoftwo
 \fi
}%
\providecommand \natexlab [1]{#1}%
\providecommand \enquote  [1]{``#1''}%
\providecommand \bibnamefont  [1]{#1}%
\providecommand \bibfnamefont [1]{#1}%
\providecommand \citenamefont [1]{#1}%
\providecommand \href@noop [0]{\@secondoftwo}%
\providecommand \href [0]{\begingroup \@sanitize@url \@href}%
\providecommand \@href[1]{\@@startlink{#1}\@@href}%
\providecommand \@@href[1]{\endgroup#1\@@endlink}%
\providecommand \@sanitize@url [0]{\catcode `\\12\catcode `\$12\catcode
  `\&12\catcode `\#12\catcode `\^12\catcode `\_12\catcode `\%12\relax}%
\providecommand \@@startlink[1]{}%
\providecommand \@@endlink[0]{}%
\providecommand \url  [0]{\begingroup\@sanitize@url \@url }%
\providecommand \@url [1]{\endgroup\@href {#1}{\urlprefix }}%
\providecommand \urlprefix  [0]{URL }%
\providecommand \Eprint [0]{\href }%
\providecommand \doibase [0]{http://dx.doi.org/}%
\providecommand \selectlanguage [0]{\@gobble}%
\providecommand \bibinfo  [0]{\@secondoftwo}%
\providecommand \bibfield  [0]{\@secondoftwo}%
\providecommand \translation [1]{[#1]}%
\providecommand \BibitemOpen [0]{}%
\providecommand \bibitemStop [0]{}%
\providecommand \bibitemNoStop [0]{.\EOS\space}%
\providecommand \EOS [0]{\spacefactor3000\relax}%
\providecommand \BibitemShut  [1]{\csname bibitem#1\endcsname}%
\let\auto@bib@innerbib\@empty
\bibitem [{\citenamefont {Willett}\ \emph {et~al.}(1993)\citenamefont
  {Willett}, \citenamefont {Ruel}, \citenamefont {West},\ and\ \citenamefont
  {Pfeiffer}}]{willett_experimental_1993}%
  \BibitemOpen
  \bibfield  {author} {\bibinfo {author} {\bibfnamefont {R.~L.}\ \bibnamefont
  {Willett}}, \bibinfo {author} {\bibfnamefont {R.~R.}\ \bibnamefont {Ruel}},
  \bibinfo {author} {\bibfnamefont {K.~W.}\ \bibnamefont {West}}, \ and\
  \bibinfo {author} {\bibfnamefont {L.~N.}\ \bibnamefont {Pfeiffer}},\
  }\bibfield  {title} {\enquote {\bibinfo {title} {Experimental demonstration
  of a {Fermi} surface at one-half filling of the lowest {Landau} level},}\
  }\href {\doibase 10.1103/PhysRevLett.71.3846} {\bibfield  {journal} {\bibinfo
   {journal} {Phys. Rev. Lett.}\ }\textbf {\bibinfo {volume} {71}},\ \bibinfo
  {pages} {3846} (\bibinfo {year} {1993})}\BibitemShut {NoStop}%
\bibitem [{\citenamefont {Kang}\ \emph {et~al.}(1993)\citenamefont {Kang},
  \citenamefont {Stormer}, \citenamefont {Pfeiffer}, \citenamefont {Baldwin},\
  and\ \citenamefont {West}}]{Kang_how_1993}%
  \BibitemOpen
  \bibfield  {author} {\bibinfo {author} {\bibfnamefont {W.}~\bibnamefont
  {Kang}}, \bibinfo {author} {\bibfnamefont {H.~L.}\ \bibnamefont {Stormer}},
  \bibinfo {author} {\bibfnamefont {L.~N.}\ \bibnamefont {Pfeiffer}}, \bibinfo
  {author} {\bibfnamefont {K.~W.}\ \bibnamefont {Baldwin}}, \ and\ \bibinfo
  {author} {\bibfnamefont {K.~W.}\ \bibnamefont {West}},\ }\bibfield  {title}
  {\enquote {\bibinfo {title} {How real are composite fermions?}}\ }\href
  {\doibase 10.1103/PhysRevLett.71.3850} {\bibfield  {journal} {\bibinfo
  {journal} {Phys. Rev. Lett.}\ }\textbf {\bibinfo {volume} {71}},\ \bibinfo
  {pages} {3850} (\bibinfo {year} {1993})}\BibitemShut {NoStop}%
\bibitem [{\citenamefont {Goldman}\ \emph {et~al.}(1994)\citenamefont
  {Goldman}, \citenamefont {Su},\ and\ \citenamefont
  {Jain}}]{Goldman_detection_1994}%
  \BibitemOpen
  \bibfield  {author} {\bibinfo {author} {\bibfnamefont {V.~J.}\ \bibnamefont
  {Goldman}}, \bibinfo {author} {\bibfnamefont {B.}~\bibnamefont {Su}}, \ and\
  \bibinfo {author} {\bibfnamefont {J.~K.}\ \bibnamefont {Jain}},\ }\bibfield
  {title} {\enquote {\bibinfo {title} {Detection of composite fermions by
  magnetic focusing},}\ }\href {\doibase 10.1103/PhysRevLett.72.2065}
  {\bibfield  {journal} {\bibinfo  {journal} {Phys. Rev. Lett.}\ }\textbf
  {\bibinfo {volume} {72}},\ \bibinfo {pages} {2065} (\bibinfo {year}
  {1994})}\BibitemShut {NoStop}%
\bibitem [{\citenamefont {Rezayi}\ and\ \citenamefont
  {Read}(1994)}]{Rezayi_fermi-liquid-like_1994}%
  \BibitemOpen
  \bibfield  {author} {\bibinfo {author} {\bibfnamefont {E.}~\bibnamefont
  {Rezayi}}\ and\ \bibinfo {author} {\bibfnamefont {N.}~\bibnamefont {Read}},\
  }\bibfield  {title} {\enquote {\bibinfo {title} {Fermi-liquid-like state in a
  half-filled {Landau} level},}\ }\href
  {https://journals.aps.org/prl/abstract/10.1103/PhysRevLett.72.900} {\bibfield
   {journal} {\bibinfo  {journal} {Phys. Rev. Lett.}\ }\textbf {\bibinfo
  {volume} {72}},\ \bibinfo {pages} {900} (\bibinfo {year} {1994})}\BibitemShut
  {NoStop}%
\bibitem [{\citenamefont {Smet}\ \emph {et~al.}(1996)\citenamefont {Smet},
  \citenamefont {Weiss}, \citenamefont {Blick}, \citenamefont {L\"utjering},
  \citenamefont {von Klitzing}, \citenamefont {Fleischmann}, \citenamefont
  {Ketzmerick}, \citenamefont {Geisel},\ and\ \citenamefont
  {Weimann}}]{Smet_magnetic_1996}%
  \BibitemOpen
  \bibfield  {author} {\bibinfo {author} {\bibfnamefont {J.~H.}\ \bibnamefont
  {Smet}}, \bibinfo {author} {\bibfnamefont {D.}~\bibnamefont {Weiss}},
  \bibinfo {author} {\bibfnamefont {R.~H.}\ \bibnamefont {Blick}}, \bibinfo
  {author} {\bibfnamefont {G.}~\bibnamefont {L\"utjering}}, \bibinfo {author}
  {\bibfnamefont {K.}~\bibnamefont {von Klitzing}}, \bibinfo {author}
  {\bibfnamefont {R.}~\bibnamefont {Fleischmann}}, \bibinfo {author}
  {\bibfnamefont {R.}~\bibnamefont {Ketzmerick}}, \bibinfo {author}
  {\bibfnamefont {T.}~\bibnamefont {Geisel}}, \ and\ \bibinfo {author}
  {\bibfnamefont {G.}~\bibnamefont {Weimann}},\ }\bibfield  {title} {\enquote
  {\bibinfo {title} {Magnetic focusing of composite fermions through arrays of
  cavities},}\ }\href {\doibase 10.1103/PhysRevLett.77.2272} {\bibfield
  {journal} {\bibinfo  {journal} {Phys. Rev. Lett.}\ }\textbf {\bibinfo
  {volume} {77}},\ \bibinfo {pages} {2272} (\bibinfo {year}
  {1996})}\BibitemShut {NoStop}%
\bibitem [{\citenamefont {Jain}(1989)}]{Jain_composite-fermion_1989}%
  \BibitemOpen
  \bibfield  {author} {\bibinfo {author} {\bibfnamefont {J.~K.}\ \bibnamefont
  {Jain}},\ }\bibfield  {title} {\enquote {\bibinfo {title} {Composite-fermion
  approach for the fractional quantum {Hall} effect},}\ }\href
  {https://journals.aps.org/prl/abstract/10.1103/PhysRevLett.63.199} {\bibfield
   {journal} {\bibinfo  {journal} {Phys. Rev. Lett.}\ }\textbf {\bibinfo
  {volume} {63}},\ \bibinfo {pages} {199} (\bibinfo {year} {1989})}\BibitemShut
  {NoStop}%
\bibitem [{\citenamefont {Jain}\ and\ \citenamefont
  {Kamilla}(1997)}]{Jain_quantitative_1997}%
  \BibitemOpen
  \bibfield  {author} {\bibinfo {author} {\bibfnamefont {J.~K.}\ \bibnamefont
  {Jain}}\ and\ \bibinfo {author} {\bibfnamefont {R.~K.}\ \bibnamefont
  {Kamilla}},\ }\bibfield  {title} {\enquote {\bibinfo {title} {Quantitative
  study of large composite-fermion systems},}\ }\href
  {https://journals.aps.org/prb/abstract/10.1103/PhysRevB.55.R4895} {\bibfield
  {journal} {\bibinfo  {journal} {Phys. Rev. B}\ }\textbf {\bibinfo {volume}
  {55}},\ \bibinfo {pages} {R4895} (\bibinfo {year} {1997})}\BibitemShut
  {NoStop}%
\bibitem [{\citenamefont {Jain}(2007)}]{Jain_composite_2007}%
  \BibitemOpen
  \bibfield  {author} {\bibinfo {author} {\bibfnamefont {J.~K.}\ \bibnamefont
  {Jain}},\ }\href {\doibase 10.1017/CBO9780511607561} {\emph {\bibinfo {title}
  {Composite {Fermions}}}}\ (\bibinfo  {publisher} {Cambridge University
  Press},\ \bibinfo {year} {2007})\BibitemShut {NoStop}%
\bibitem [{\citenamefont {Halperin}\ \emph {et~al.}(1993)\citenamefont
  {Halperin}, \citenamefont {Lee},\ and\ \citenamefont
  {Read}}]{halperin_theory_1993}%
  \BibitemOpen
  \bibfield  {author} {\bibinfo {author} {\bibfnamefont {B.~I.}\ \bibnamefont
  {Halperin}}, \bibinfo {author} {\bibfnamefont {P.~A.}\ \bibnamefont {Lee}}, \
  and\ \bibinfo {author} {\bibfnamefont {N.}~\bibnamefont {Read}},\ }\bibfield
  {title} {\enquote {\bibinfo {title} {Theory of the half-filled {Landau}
  level},}\ }\href
  {https://journals.aps.org/prb/abstract/10.1103/PhysRevB.47.7312} {\bibfield
  {journal} {\bibinfo  {journal} {Phys. Rev. B}\ }\textbf {\bibinfo {volume}
  {47}},\ \bibinfo {pages} {7312} (\bibinfo {year} {1993})}\BibitemShut
  {NoStop}%
\bibitem [{\citenamefont {Willett}\ \emph {et~al.}(1987)\citenamefont
  {Willett}, \citenamefont {Eisenstein}, \citenamefont {St\"ormer},
  \citenamefont {Tsui}, \citenamefont {Gossard},\ and\ \citenamefont
  {English}}]{Willett_observation_1987}%
  \BibitemOpen
  \bibfield  {author} {\bibinfo {author} {\bibfnamefont {R.}~\bibnamefont
  {Willett}}, \bibinfo {author} {\bibfnamefont {J.~P.}\ \bibnamefont
  {Eisenstein}}, \bibinfo {author} {\bibfnamefont {H.~L.}\ \bibnamefont
  {St\"ormer}}, \bibinfo {author} {\bibfnamefont {D.~C.}\ \bibnamefont {Tsui}},
  \bibinfo {author} {\bibfnamefont {A.~C.}\ \bibnamefont {Gossard}}, \ and\
  \bibinfo {author} {\bibfnamefont {J.~H.}\ \bibnamefont {English}},\
  }\bibfield  {title} {\enquote {\bibinfo {title} {Observation of an
  even-denominator quantum number in the fractional quantum {Hall} effect},}\
  }\href {\doibase 10.1103/PhysRevLett.59.1776} {\bibfield  {journal} {\bibinfo
   {journal} {Phys. Rev. Lett.}\ }\textbf {\bibinfo {volume} {59}},\ \bibinfo
  {pages} {1776} (\bibinfo {year} {1987})}\BibitemShut {NoStop}%
\bibitem [{\citenamefont {Morf}(1998)}]{Morf_transition_1998}%
  \BibitemOpen
  \bibfield  {author} {\bibinfo {author} {\bibfnamefont {R.~H.}\ \bibnamefont
  {Morf}},\ }\bibfield  {title} {\enquote {\bibinfo {title} {Transition from
  quantum {Hall} to compressible states in the second {Landau} level: new light
  on the $\nu=$~5/2 enigma},}\ }\href
  {https://journals.aps.org/prl/abstract/10.1103/PhysRevLett.80.1505}
  {\bibfield  {journal} {\bibinfo  {journal} {Phys. Rev. Lett.}\ }\textbf
  {\bibinfo {volume} {80}},\ \bibinfo {pages} {1505} (\bibinfo {year}
  {1998})}\BibitemShut {NoStop}%
\bibitem [{\citenamefont {Rezayi}\ and\ \citenamefont
  {Haldane}(2000)}]{Rezayi_incompressible_2000}%
  \BibitemOpen
  \bibfield  {author} {\bibinfo {author} {\bibfnamefont {E.~H.}\ \bibnamefont
  {Rezayi}}\ and\ \bibinfo {author} {\bibfnamefont {F.~D.~M.}\ \bibnamefont
  {Haldane}},\ }\bibfield  {title} {\enquote {\bibinfo {title} {Incompressible
  paired {Hall} state, stripe order, and the composite fermion liquid phase in
  half-filled {Landau} levels},}\ }\href
  {https://journals.aps.org/prl/abstract/10.1103/PhysRevLett.84.4685}
  {\bibfield  {journal} {\bibinfo  {journal} {Phys. Rev. Lett.}\ }\textbf
  {\bibinfo {volume} {84}},\ \bibinfo {pages} {4685} (\bibinfo {year}
  {2000})}\BibitemShut {NoStop}%
\bibitem [{\citenamefont {W\'ojs}\ \emph {et~al.}(2010)\citenamefont {W\'ojs},
  \citenamefont {Toke},\ and\ \citenamefont {Jain}}]{Wojs_landau_level_2010}%
  \BibitemOpen
  \bibfield  {author} {\bibinfo {author} {\bibfnamefont {A.}~\bibnamefont
  {W\'ojs}}, \bibinfo {author} {\bibfnamefont {C.}~\bibnamefont {Toke}}, \ and\
  \bibinfo {author} {\bibfnamefont {J.~K.}\ \bibnamefont {Jain}},\ }\bibfield
  {title} {\enquote {\bibinfo {title} {Landau-level mixing and the emergence of
  {Pfaffian} excitations for the $5/2$ fractional quantum {Hall} effect},}\
  }\href {\doibase 10.1103/PhysRevLett.105.096802} {\bibfield  {journal}
  {\bibinfo  {journal} {Phys. Rev. Lett.}\ }\textbf {\bibinfo {volume} {105}},\
  \bibinfo {pages} {096802} (\bibinfo {year} {2010})}\BibitemShut {NoStop}%
\bibitem [{\citenamefont {Storni}\ \emph {et~al.}(2010)\citenamefont {Storni},
  \citenamefont {Morf},\ and\ \citenamefont
  {Das~Sarma}}]{Storni_fractional_2010}%
  \BibitemOpen
  \bibfield  {author} {\bibinfo {author} {\bibfnamefont {M.}~\bibnamefont
  {Storni}}, \bibinfo {author} {\bibfnamefont {R.~H.}\ \bibnamefont {Morf}}, \
  and\ \bibinfo {author} {\bibfnamefont {S.}~\bibnamefont {Das~Sarma}},\
  }\bibfield  {title} {\enquote {\bibinfo {title} {Fractional quantum {Hall}
  state at $\nu=$~5/2 and the {Moore}-{Read} {Pfaffian}},}\ }\href {\doibase
  10.1103/PhysRevLett.104.076803} {\bibfield  {journal} {\bibinfo  {journal}
  {Phys. Rev. Lett.}\ }\textbf {\bibinfo {volume} {104}},\ \bibinfo {pages}
  {076803} (\bibinfo {year} {2010})}\BibitemShut {NoStop}%
\bibitem [{\citenamefont {Feiguin}\ \emph {et~al.}(2008)\citenamefont
  {Feiguin}, \citenamefont {Rezayi}, \citenamefont {Nayak},\ and\ \citenamefont
  {Das~Sarma}}]{feiguin_density_2008}%
  \BibitemOpen
  \bibfield  {author} {\bibinfo {author} {\bibfnamefont {A.~E.}\ \bibnamefont
  {Feiguin}}, \bibinfo {author} {\bibfnamefont {E.}~\bibnamefont {Rezayi}},
  \bibinfo {author} {\bibfnamefont {C.}~\bibnamefont {Nayak}}, \ and\ \bibinfo
  {author} {\bibfnamefont {S.}~\bibnamefont {Das~Sarma}},\ }\bibfield  {title}
  {\enquote {\bibinfo {title} {Density matrix renormalization group study of
  incompressible fractional quantum {Hall} states},}\ }\href {\doibase
  10.1103/PhysRevLett.100.166803} {\bibfield  {journal} {\bibinfo  {journal}
  {Phys. Rev. Lett.}\ }\textbf {\bibinfo {volume} {100}},\ \bibinfo {pages}
  {166803} (\bibinfo {year} {2008})}\BibitemShut {NoStop}%
\bibitem [{\citenamefont {Feiguin}\ \emph {et~al.}(2009)\citenamefont
  {Feiguin}, \citenamefont {Rezayi}, \citenamefont {Yang}, \citenamefont
  {Nayak},\ and\ \citenamefont {Das~Sarma}}]{Feiguin_spin_2009}%
  \BibitemOpen
  \bibfield  {author} {\bibinfo {author} {\bibfnamefont {A.~E.}\ \bibnamefont
  {Feiguin}}, \bibinfo {author} {\bibfnamefont {E.}~\bibnamefont {Rezayi}},
  \bibinfo {author} {\bibfnamefont {K.}~\bibnamefont {Yang}}, \bibinfo {author}
  {\bibfnamefont {C.}~\bibnamefont {Nayak}}, \ and\ \bibinfo {author}
  {\bibfnamefont {S.}~\bibnamefont {Das~Sarma}},\ }\bibfield  {title} {\enquote
  {\bibinfo {title} {Spin polarization of the $\ensuremath{\nu}=5/2$ quantum
  {Hall} state},}\ }\href {\doibase 10.1103/PhysRevB.79.115322} {\bibfield
  {journal} {\bibinfo  {journal} {Phys. Rev. B}\ }\textbf {\bibinfo {volume}
  {79}},\ \bibinfo {pages} {115322} (\bibinfo {year} {2009})}\BibitemShut
  {NoStop}%
\bibitem [{\citenamefont {Peterson}\ \emph {et~al.}(2008)\citenamefont
  {Peterson}, \citenamefont {Jolicoeur},\ and\ \citenamefont
  {Das~Sarma}}]{Peterson_Finite_Layer_Thickness_2008}%
  \BibitemOpen
  \bibfield  {author} {\bibinfo {author} {\bibfnamefont {M.~R.}\ \bibnamefont
  {Peterson}}, \bibinfo {author} {\bibfnamefont {Th.}\ \bibnamefont
  {Jolicoeur}}, \ and\ \bibinfo {author} {\bibfnamefont {S.}~\bibnamefont
  {Das~Sarma}},\ }\bibfield  {title} {\enquote {\bibinfo {title} {Finite-layer
  thickness stabilizes the {Pfaffian} state for the 5/2 fractional quantum
  {Hall} effect: Wave function overlap and topological degeneracy},}\ }\href
  {\doibase 10.1103/PhysRevLett.101.016807} {\bibfield  {journal} {\bibinfo
  {journal} {Phys. Rev. Lett.}\ }\textbf {\bibinfo {volume} {101}},\ \bibinfo
  {pages} {016807} (\bibinfo {year} {2008})}\BibitemShut {NoStop}%
\bibitem [{\citenamefont {Rezayi}\ and\ \citenamefont
  {Simon}(2011)}]{Rezayi_breaking_2011}%
  \BibitemOpen
  \bibfield  {author} {\bibinfo {author} {\bibfnamefont {E.~H.}\ \bibnamefont
  {Rezayi}}\ and\ \bibinfo {author} {\bibfnamefont {S.~H.}\ \bibnamefont
  {Simon}},\ }\bibfield  {title} {\enquote {\bibinfo {title} {Breaking of
  particle-hole symmetry by {Landau} level mixing in the $\nu = 5/2$ quantized
  {Hall} state},}\ }\href {\doibase 10.1103/PhysRevLett.106.116801} {\bibfield
  {journal} {\bibinfo  {journal} {Phys. Rev. Lett.}\ }\textbf {\bibinfo
  {volume} {106}},\ \bibinfo {pages} {116801} (\bibinfo {year}
  {2011})}\BibitemShut {NoStop}%
\bibitem [{\citenamefont {Pakrouski}\ \emph {et~al.}(2015)\citenamefont
  {Pakrouski}, \citenamefont {Peterson}, \citenamefont {Jolicoeur},
  \citenamefont {Scarola}, \citenamefont {Nayak},\ and\ \citenamefont
  {Troyer}}]{Pakrouski_phase_diagram_2015}%
  \BibitemOpen
  \bibfield  {author} {\bibinfo {author} {\bibfnamefont {K.}~\bibnamefont
  {Pakrouski}}, \bibinfo {author} {\bibfnamefont {M.~R.}\ \bibnamefont
  {Peterson}}, \bibinfo {author} {\bibfnamefont {Th.}\ \bibnamefont
  {Jolicoeur}}, \bibinfo {author} {\bibfnamefont {V.~W.}\ \bibnamefont
  {Scarola}}, \bibinfo {author} {\bibfnamefont {C.}~\bibnamefont {Nayak}}, \
  and\ \bibinfo {author} {\bibfnamefont {M.}~\bibnamefont {Troyer}},\
  }\bibfield  {title} {\enquote {\bibinfo {title} {Phase diagram of the
  $\nu=5/2$ fractional quantum {Hall} effect: Effects of {Landau}-level mixing
  and nonzero width},}\ }\href {\doibase 10.1103/PhysRevX.5.021004} {\bibfield
  {journal} {\bibinfo  {journal} {Phys. Rev. X}\ }\textbf {\bibinfo {volume}
  {5}},\ \bibinfo {pages} {021004} (\bibinfo {year} {2015})}\BibitemShut
  {NoStop}%
\bibitem [{\citenamefont {Halperin}(1983)}]{Halperin_QH_1983}%
  \BibitemOpen
  \bibfield  {author} {\bibinfo {author} {\bibfnamefont {B.I.}\ \bibnamefont
  {Halperin}},\ }\bibfield  {title} {\enquote {\bibinfo {title} {Theory of the
  quantized {Hall} conductance},}\ }\href {\doibase
  http://doi.org/10.5169/seals-115362} {\bibfield  {journal} {\bibinfo
  {journal} {Helv. Phys. Acta {\bf 56}}\ ,\ \bibinfo {pages} {75}} (\bibinfo
  {year} {1983})}\BibitemShut {NoStop}%
\bibitem [{\citenamefont {Moore}\ and\ \citenamefont
  {Read}(1991)}]{Moore_nonabelions_1991}%
  \BibitemOpen
  \bibfield  {author} {\bibinfo {author} {\bibfnamefont {G.}~\bibnamefont
  {Moore}}\ and\ \bibinfo {author} {\bibfnamefont {N.}~\bibnamefont {Read}},\
  }\bibfield  {title} {\enquote {\bibinfo {title} {Nonabelions in the
  fractional quantum {Hall} effect},}\ }\href
  {http://www.sciencedirect.com/science/article/pii/055032139190407O}
  {\bibfield  {journal} {\bibinfo  {journal} {Nucl. Phys. B}\ }\textbf
  {\bibinfo {volume} {360}},\ \bibinfo {pages} {362} (\bibinfo {year}
  {1991})}\BibitemShut {NoStop}%
\bibitem [{\citenamefont {Greiter}\ \emph {et~al.}(1991)\citenamefont
  {Greiter}, \citenamefont {Wen},\ and\ \citenamefont
  {Wilczek}}]{Greiter_half_filled_1991}%
  \BibitemOpen
  \bibfield  {author} {\bibinfo {author} {\bibfnamefont {M.}~\bibnamefont
  {Greiter}}, \bibinfo {author} {\bibfnamefont {X.~G.}\ \bibnamefont {Wen}}, \
  and\ \bibinfo {author} {\bibfnamefont {F.}~\bibnamefont {Wilczek}},\
  }\bibfield  {title} {\enquote {\bibinfo {title} {Paired {Hall} state at half
  filling},}\ }\href {\doibase 10.1103/PhysRevLett.66.3205} {\bibfield
  {journal} {\bibinfo  {journal} {Phys. Rev. Lett.}\ }\textbf {\bibinfo
  {volume} {66}},\ \bibinfo {pages} {3205} (\bibinfo {year}
  {1991})}\BibitemShut {NoStop}%
\bibitem [{\citenamefont {Haldane}\ and\ \citenamefont
  {Rezayi}(1988)}]{Haldane_spin_singlet_1988}%
  \BibitemOpen
  \bibfield  {author} {\bibinfo {author} {\bibfnamefont {F.~D.~M.}\
  \bibnamefont {Haldane}}\ and\ \bibinfo {author} {\bibfnamefont {E.~H.}\
  \bibnamefont {Rezayi}},\ }\bibfield  {title} {\enquote {\bibinfo {title}
  {Spin-singlet wave function for the half-integral quantum {Hall} effect},}\
  }\href {\doibase 10.1103/PhysRevLett.60.956} {\bibfield  {journal} {\bibinfo
  {journal} {Phys. Rev. Lett.}\ }\textbf {\bibinfo {volume} {60}},\ \bibinfo
  {pages} {956} (\bibinfo {year} {1988})}\BibitemShut {NoStop}%
\bibitem [{\citenamefont {Read}\ and\ \citenamefont
  {Green}(2000)}]{Read_paired_2000}%
  \BibitemOpen
  \bibfield  {author} {\bibinfo {author} {\bibfnamefont {N.}~\bibnamefont
  {Read}}\ and\ \bibinfo {author} {\bibfnamefont {D.}~\bibnamefont {Green}},\
  }\bibfield  {title} {\enquote {\bibinfo {title} {Paired states of fermions in
  two dimensions with breaking of parity and time-reversal symmetries and the
  fractional quantum {Hall} effect},}\ }\href
  {https://journals.aps.org/prb/abstract/10.1103/PhysRevB.61.10267} {\bibfield
  {journal} {\bibinfo  {journal} {Phys. Rev. B}\ }\textbf {\bibinfo {volume}
  {61}},\ \bibinfo {pages} {10267} (\bibinfo {year} {2000})}\BibitemShut
  {NoStop}%
\bibitem [{\citenamefont {Levin}\ \emph {et~al.}(2007)\citenamefont {Levin},
  \citenamefont {Halperin},\ and\ \citenamefont
  {Rosenow}}]{Levin_particle-hole_2007}%
  \BibitemOpen
  \bibfield  {author} {\bibinfo {author} {\bibfnamefont {M.}~\bibnamefont
  {Levin}}, \bibinfo {author} {\bibfnamefont {B.~I.}\ \bibnamefont {Halperin}},
  \ and\ \bibinfo {author} {\bibfnamefont {B.}~\bibnamefont {Rosenow}},\
  }\bibfield  {title} {\enquote {\bibinfo {title} {Particle-hole symmetry and
  the {Pfaffian} state},}\ }\href {\doibase 10.1103/PhysRevLett.99.236806}
  {\bibfield  {journal} {\bibinfo  {journal} {Phys. Rev. Lett.}\ }\textbf
  {\bibinfo {volume} {99}},\ \bibinfo {pages} {236806} (\bibinfo {year}
  {2007})}\BibitemShut {NoStop}%
\bibitem [{\citenamefont {Lee}\ \emph {et~al.}(2007)\citenamefont {Lee},
  \citenamefont {Ryu}, \citenamefont {Nayak},\ and\ \citenamefont
  {Fisher}}]{Lee_particle-hole_2007}%
  \BibitemOpen
  \bibfield  {author} {\bibinfo {author} {\bibfnamefont {S.~S.}\ \bibnamefont
  {Lee}}, \bibinfo {author} {\bibfnamefont {S.}~\bibnamefont {Ryu}}, \bibinfo
  {author} {\bibfnamefont {C.}~\bibnamefont {Nayak}}, \ and\ \bibinfo {author}
  {\bibfnamefont {M.~P.~A.}\ \bibnamefont {Fisher}},\ }\bibfield  {title}
  {\enquote {\bibinfo {title} {Particle-hole symmetry and the $\nu=$~5/2
  quantum {Hall} state},}\ }\href {\doibase 10.1103/PhysRevLett.99.236807}
  {\bibfield  {journal} {\bibinfo  {journal} {Phys. Rev. Lett.}\ }\textbf
  {\bibinfo {volume} {99}},\ \bibinfo {pages} {236807} (\bibinfo {year}
  {2007})}\BibitemShut {NoStop}%
\bibitem [{\citenamefont {Girvin}(1984)}]{Girvin_PHS_1984}%
  \BibitemOpen
  \bibfield  {author} {\bibinfo {author} {\bibfnamefont {S.~M.}\ \bibnamefont
  {Girvin}},\ }\bibfield  {title} {\enquote {\bibinfo {title} {Particle-hole
  symmetry in the anomalous quantum {Hall} effect},}\ }\href {\doibase
  10.1103/PhysRevB.29.6012} {\bibfield  {journal} {\bibinfo  {journal} {Phys.
  Rev. B}\ }\textbf {\bibinfo {volume} {29}},\ \bibinfo {pages} {6012}
  (\bibinfo {year} {1984})}\BibitemShut {NoStop}%
\bibitem [{\citenamefont {Son}(2015)}]{Son_is_2015}%
  \BibitemOpen
  \bibfield  {author} {\bibinfo {author} {\bibfnamefont {D.~T.}\ \bibnamefont
  {Son}},\ }\bibfield  {title} {\enquote {\bibinfo {title} {Is the composite
  {Fermion} a {Dirac} particle?}}\ }\href {\doibase 10.1103/PhysRevX.5.031027}
  {\bibfield  {journal} {\bibinfo  {journal} {Phys. Rev. X}\ }\textbf {\bibinfo
  {volume} {5}},\ \bibinfo {pages} {031027} (\bibinfo {year}
  {2015})}\BibitemShut {NoStop}%
\bibitem [{\citenamefont {Metlitski}\ and\ \citenamefont
  {Vishwanath}(2016)}]{Metlitski_particle_vortex_2016}%
  \BibitemOpen
  \bibfield  {author} {\bibinfo {author} {\bibfnamefont {M.~A.}\ \bibnamefont
  {Metlitski}}\ and\ \bibinfo {author} {\bibfnamefont {A.}~\bibnamefont
  {Vishwanath}},\ }\bibfield  {title} {\enquote {\bibinfo {title}
  {Particle-vortex duality of two-dimensional {Dirac} fermion from
  electric-magnetic duality of three-dimensional topological insulators},}\
  }\href {\doibase 10.1103/PhysRevB.93.245151} {\bibfield  {journal} {\bibinfo
  {journal} {Phys. Rev. B}\ }\textbf {\bibinfo {volume} {93}},\ \bibinfo
  {pages} {245151} (\bibinfo {year} {2016})}\BibitemShut {NoStop}%
\bibitem [{\citenamefont {Wang}\ and\ \citenamefont
  {Senthil}(2016)}]{Wang_half_filled_2016}%
  \BibitemOpen
  \bibfield  {author} {\bibinfo {author} {\bibfnamefont {C.}~\bibnamefont
  {Wang}}\ and\ \bibinfo {author} {\bibfnamefont {T.}~\bibnamefont {Senthil}},\
  }\bibfield  {title} {\enquote {\bibinfo {title} {Half-filled {Landau} level,
  topological insulator surfaces, and three-dimensional quantum spin
  liquids},}\ }\href {\doibase 10.1103/PhysRevB.93.085110} {\bibfield
  {journal} {\bibinfo  {journal} {Phys. Rev. B}\ }\textbf {\bibinfo {volume}
  {93}},\ \bibinfo {pages} {085110} (\bibinfo {year} {2016})}\BibitemShut
  {NoStop}%
\bibitem [{\citenamefont {Geraedts}\ \emph {et~al.}(2016)\citenamefont
  {Geraedts}, \citenamefont {Zaletel}, \citenamefont {Mong}, \citenamefont
  {Metlitski}, \citenamefont {Vishwanath},\ and\ \citenamefont
  {Motrunich}}]{Geraedts_half_filled_2016}%
  \BibitemOpen
  \bibfield  {author} {\bibinfo {author} {\bibfnamefont {S.~D.}\ \bibnamefont
  {Geraedts}}, \bibinfo {author} {\bibfnamefont {M.~P.}\ \bibnamefont
  {Zaletel}}, \bibinfo {author} {\bibfnamefont {R.~S.~K.}\ \bibnamefont
  {Mong}}, \bibinfo {author} {\bibfnamefont {M.~A.}\ \bibnamefont {Metlitski}},
  \bibinfo {author} {\bibfnamefont {A.}~\bibnamefont {Vishwanath}}, \ and\
  \bibinfo {author} {\bibfnamefont {O.~I.}\ \bibnamefont {Motrunich}},\
  }\bibfield  {title} {\enquote {\bibinfo {title} {The half-filled {Landau}
  level: {The} case for {Dirac} composite fermions},}\ }\href {\doibase
  10.1126/science.aad4302} {\bibfield  {journal} {\bibinfo  {journal}
  {Science}\ }\textbf {\bibinfo {volume} {352}},\ \bibinfo {pages} {197}
  (\bibinfo {year} {2016})}\BibitemShut {NoStop}%
\bibitem [{\citenamefont {Murthy}\ and\ \citenamefont
  {Shankar}(2016)}]{Murthy_half_filled_2016}%
  \BibitemOpen
  \bibfield  {author} {\bibinfo {author} {\bibfnamefont {G.}~\bibnamefont
  {Murthy}}\ and\ \bibinfo {author} {\bibfnamefont {R.}~\bibnamefont
  {Shankar}},\ }\bibfield  {title} {\enquote {\bibinfo {title} {$\nu=$~1/2
  {Landau} level: Half-empty versus half-full},}\ }\href {\doibase
  10.1103/PhysRevB.93.085405} {\bibfield  {journal} {\bibinfo  {journal} {Phys.
  Rev. B}\ }\textbf {\bibinfo {volume} {93}},\ \bibinfo {pages} {085405}
  (\bibinfo {year} {2016})}\BibitemShut {NoStop}%
\bibitem [{\citenamefont {Kachru}\ \emph {et~al.}(2015)\citenamefont {Kachru},
  \citenamefont {Mulligan}, \citenamefont {Torroba},\ and\ \citenamefont
  {Wang}}]{Kachru_half_filled_2016}%
  \BibitemOpen
  \bibfield  {author} {\bibinfo {author} {\bibfnamefont {S.}~\bibnamefont
  {Kachru}}, \bibinfo {author} {\bibfnamefont {M.}~\bibnamefont {Mulligan}},
  \bibinfo {author} {\bibfnamefont {G.}~\bibnamefont {Torroba}}, \ and\
  \bibinfo {author} {\bibfnamefont {H.}~\bibnamefont {Wang}},\ }\bibfield
  {title} {\enquote {\bibinfo {title} {Mirror symmetry and the half-filled
  {Landau} level},}\ }\href {\doibase 10.1103/PhysRevB.92.235105} {\bibfield
  {journal} {\bibinfo  {journal} {Phys. Rev. B}\ }\textbf {\bibinfo {volume}
  {92}},\ \bibinfo {pages} {235105} (\bibinfo {year} {2015})}\BibitemShut
  {NoStop}%
\bibitem [{\citenamefont {Mross}\ \emph {et~al.}(2016)\citenamefont {Mross},
  \citenamefont {Alicea},\ and\ \citenamefont
  {Motrunich}}]{Mross_explicit_duality_2016}%
  \BibitemOpen
  \bibfield  {author} {\bibinfo {author} {\bibfnamefont {D.~F.}\ \bibnamefont
  {Mross}}, \bibinfo {author} {\bibfnamefont {J.}~\bibnamefont {Alicea}}, \
  and\ \bibinfo {author} {\bibfnamefont {O.~I.}\ \bibnamefont {Motrunich}},\
  }\bibfield  {title} {\enquote {\bibinfo {title} {Explicit derivation of
  duality between a free {Dirac} cone and quantum electrodynamics in ($2+1$)
  dimensions},}\ }\href {\doibase 10.1103/PhysRevLett.117.016802} {\bibfield
  {journal} {\bibinfo  {journal} {Phys. Rev. Lett.}\ }\textbf {\bibinfo
  {volume} {117}},\ \bibinfo {pages} {016802} (\bibinfo {year}
  {2016})}\BibitemShut {NoStop}%
\bibitem [{\citenamefont {Mulligan}\ \emph {et~al.}(2016)\citenamefont
  {Mulligan}, \citenamefont {Raghu},\ and\ \citenamefont
  {Fisher}}]{Mulligan_emergent_ph_2016}%
  \BibitemOpen
  \bibfield  {author} {\bibinfo {author} {\bibfnamefont {M.}~\bibnamefont
  {Mulligan}}, \bibinfo {author} {\bibfnamefont {S.}~\bibnamefont {Raghu}}, \
  and\ \bibinfo {author} {\bibfnamefont {M.~P.~A.}\ \bibnamefont {Fisher}},\
  }\bibfield  {title} {\enquote {\bibinfo {title} {Emergent particle-hole
  symmetry in the half-filled {Landau} level},}\ }\href {\doibase
  10.1103/PhysRevB.94.075101} {\bibfield  {journal} {\bibinfo  {journal} {Phys.
  Rev. B}\ }\textbf {\bibinfo {volume} {94}},\ \bibinfo {pages} {075101}
  (\bibinfo {year} {2016})}\BibitemShut {NoStop}%
\bibitem [{\citenamefont {Balram}\ and\ \citenamefont
  {Jain}(2016)}]{Balram_nature_2016}%
  \BibitemOpen
  \bibfield  {author} {\bibinfo {author} {\bibfnamefont {A.~C.}\ \bibnamefont
  {Balram}}\ and\ \bibinfo {author} {\bibfnamefont {J.~K.}\ \bibnamefont
  {Jain}},\ }\bibfield  {title} {\enquote {\bibinfo {title} {Nature of
  composite fermions and the role of particle-hole symmetry: {A} microscopic
  account},}\ }\href
  {https://journals.aps.org/prb/abstract/10.1103/PhysRevB.93.235152} {\bibfield
   {journal} {\bibinfo  {journal} {Phys. Rev. B}\ }\textbf {\bibinfo {volume}
  {93}},\ \bibinfo {pages} {235152} (\bibinfo {year} {2016})}\BibitemShut
  {NoStop}%
\bibitem [{\citenamefont {{Yang}}(2017)}]{Yang_dirac_2017}%
  \BibitemOpen
  \bibfield  {author} {\bibinfo {author} {\bibfnamefont {J.}~\bibnamefont
  {{Yang}}},\ }\bibfield  {title} {\enquote {\bibinfo {title} {Dirac composite
  fermion - {A} particle-hole spinor},}\ }\href
  {https://arxiv.org/abs/1711.08520} {\bibfield  {journal} {\bibinfo  {journal}
  {arXiv:1711.08520}\ } (\bibinfo {year} {2017})}\BibitemShut {NoStop}%
\bibitem [{\citenamefont {Fremling}\ \emph {et~al.}(2018)\citenamefont
  {Fremling}, \citenamefont {Moran}, \citenamefont {Slingerland},\ and\
  \citenamefont {Simon}}]{Fremling_trial_2018}%
  \BibitemOpen
  \bibfield  {author} {\bibinfo {author} {\bibfnamefont {M.}~\bibnamefont
  {Fremling}}, \bibinfo {author} {\bibfnamefont {N.}~\bibnamefont {Moran}},
  \bibinfo {author} {\bibfnamefont {J.~K.}\ \bibnamefont {Slingerland}}, \ and\
  \bibinfo {author} {\bibfnamefont {S.~H.}\ \bibnamefont {Simon}},\ }\bibfield
  {title} {\enquote {\bibinfo {title} {Trial wave functions for a composite
  {Fermi} liquid on a torus},}\ }\href {\doibase 10.1103/PhysRevB.97.035149}
  {\bibfield  {journal} {\bibinfo  {journal} {Phys. Rev. B}\ }\textbf {\bibinfo
  {volume} {97}},\ \bibinfo {pages} {035149} (\bibinfo {year}
  {2018})}\BibitemShut {NoStop}%
\bibitem [{\citenamefont {Chen}\ \emph {et~al.}(2014)\citenamefont {Chen},
  \citenamefont {Fidkowski},\ and\ \citenamefont
  {Vishwanath}}]{chen_symmetry_2014}%
  \BibitemOpen
  \bibfield  {author} {\bibinfo {author} {\bibfnamefont {X.}~\bibnamefont
  {Chen}}, \bibinfo {author} {\bibfnamefont {L.}~\bibnamefont {Fidkowski}}, \
  and\ \bibinfo {author} {\bibfnamefont {A.}~\bibnamefont {Vishwanath}},\
  }\bibfield  {title} {\enquote {\bibinfo {title} {Symmetry enforced
  non-{Abelian} topological order at the surface of a topological insulator},}\
  }\href {\doibase 10.1103/PhysRevB.89.165132} {\bibfield  {journal} {\bibinfo
  {journal} {Phys. Rev. B}\ }\textbf {\bibinfo {volume} {89}},\ \bibinfo
  {pages} {165132} (\bibinfo {year} {2014})}\BibitemShut {NoStop}%
\bibitem [{\citenamefont {Bonderson}\ \emph {et~al.}(2013)\citenamefont
  {Bonderson}, \citenamefont {Nayak},\ and\ \citenamefont
  {Qi}}]{Bonderson_time-reversal_2013}%
  \BibitemOpen
  \bibfield  {author} {\bibinfo {author} {\bibfnamefont {P.}~\bibnamefont
  {Bonderson}}, \bibinfo {author} {\bibfnamefont {C.}~\bibnamefont {Nayak}}, \
  and\ \bibinfo {author} {\bibfnamefont {X.~L.}\ \bibnamefont {Qi}},\
  }\bibfield  {title} {\enquote {\bibinfo {title} {A time-reversal invariant
  topological phase at the surface of a 3{D} topological insulator},}\ }\href
  {\doibase 10.1088/1742-5468/2013/09/P09016} {\bibfield  {journal} {\bibinfo
  {journal} {J. Stat. Mech.}\ }\textbf {\bibinfo {volume} {2013}},\ \bibinfo
  {pages} {P09016} (\bibinfo {year} {2013})}\BibitemShut {NoStop}%
\bibitem [{\citenamefont {Metlitski}\ \emph {et~al.}(2015)\citenamefont
  {Metlitski}, \citenamefont {Kane},\ and\ \citenamefont
  {Fisher}}]{Metlitski_symmetry_respecting_2015}%
  \BibitemOpen
  \bibfield  {author} {\bibinfo {author} {\bibfnamefont {M.~A.}\ \bibnamefont
  {Metlitski}}, \bibinfo {author} {\bibfnamefont {C.~L.}\ \bibnamefont {Kane}},
  \ and\ \bibinfo {author} {\bibfnamefont {M.~P.~A.}\ \bibnamefont {Fisher}},\
  }\bibfield  {title} {\enquote {\bibinfo {title} {Symmetry-respecting
  topologically ordered surface phase of three-dimensional electron topological
  insulators},}\ }\href {\doibase 10.1103/PhysRevB.92.125111} {\bibfield
  {journal} {\bibinfo  {journal} {Phys. Rev. B}\ }\textbf {\bibinfo {volume}
  {92}},\ \bibinfo {pages} {125111} (\bibinfo {year} {2015})}\BibitemShut
  {NoStop}%
\bibitem [{\citenamefont {Wang}\ \emph {et~al.}(2013)\citenamefont {Wang},
  \citenamefont {Potter},\ and\ \citenamefont {Senthil}}]{Wang_gapped_2013}%
  \BibitemOpen
  \bibfield  {author} {\bibinfo {author} {\bibfnamefont {C.}~\bibnamefont
  {Wang}}, \bibinfo {author} {\bibfnamefont {A.~C.}\ \bibnamefont {Potter}}, \
  and\ \bibinfo {author} {\bibfnamefont {T.}~\bibnamefont {Senthil}},\
  }\bibfield  {title} {\enquote {\bibinfo {title} {Gapped symmetry preserving
  surface state for the electron topological insulator},}\ }\href {\doibase
  10.1103/PhysRevB.88.115137} {\bibfield  {journal} {\bibinfo  {journal} {Phys.
  Rev. B}\ }\textbf {\bibinfo {volume} {88}},\ \bibinfo {pages} {115137}
  (\bibinfo {year} {2013})}\BibitemShut {NoStop}%
\bibitem [{\citenamefont {Mross}\ \emph {et~al.}(2015)\citenamefont {Mross},
  \citenamefont {Essin},\ and\ \citenamefont
  {Alicea}}]{Mross_Composite_Dirac_Liquids_2015}%
  \BibitemOpen
  \bibfield  {author} {\bibinfo {author} {\bibfnamefont {D.~F.}\ \bibnamefont
  {Mross}}, \bibinfo {author} {\bibfnamefont {A.}~\bibnamefont {Essin}}, \ and\
  \bibinfo {author} {\bibfnamefont {J.}~\bibnamefont {Alicea}},\ }\bibfield
  {title} {\enquote {\bibinfo {title} {Composite {Dirac} liquids: Parent states
  for symmetric surface topological order},}\ }\href {\doibase
  10.1103/PhysRevX.5.011011} {\bibfield  {journal} {\bibinfo  {journal} {Phys.
  Rev. X}\ }\textbf {\bibinfo {volume} {5}},\ \bibinfo {pages} {011011}
  (\bibinfo {year} {2015})}\BibitemShut {NoStop}%
\bibitem [{\citenamefont {{Dolev}}\ \emph {et~al.}(2008)\citenamefont
  {{Dolev}}, \citenamefont {{Heiblum}}, \citenamefont {{Umansky}},
  \citenamefont {{Stern}},\ and\ \citenamefont
  {{Mahalu}}}]{Dolev_observation_2008}%
  \BibitemOpen
  \bibfield  {author} {\bibinfo {author} {\bibfnamefont {M.}~\bibnamefont
  {{Dolev}}}, \bibinfo {author} {\bibfnamefont {M.}~\bibnamefont {{Heiblum}}},
  \bibinfo {author} {\bibfnamefont {V.}~\bibnamefont {{Umansky}}}, \bibinfo
  {author} {\bibfnamefont {A.}~\bibnamefont {{Stern}}}, \ and\ \bibinfo
  {author} {\bibfnamefont {D.}~\bibnamefont {{Mahalu}}},\ }\bibfield  {title}
  {\enquote {\bibinfo {title} {{Observation of a quarter of an electron charge
  at the $\nu=$~5/2 quantum {Hall} state}},}\ }\href {\doibase
  10.1038/nature06855} {\bibfield  {journal} {\bibinfo  {journal} {Nature
  (London)}\ }\textbf {\bibinfo {volume} {452}},\ \bibinfo {pages} {829}
  (\bibinfo {year} {2008})}\BibitemShut {NoStop}%
\bibitem [{\citenamefont {Banerjee}\ \emph {et~al.}(2018)\citenamefont
  {Banerjee}, \citenamefont {Heiblum}, \citenamefont {Umansky}, \citenamefont
  {Feldman}, \citenamefont {Oreg},\ and\ \citenamefont
  {Stern}}]{Banerjee_observation_2018}%
  \BibitemOpen
  \bibfield  {author} {\bibinfo {author} {\bibfnamefont {M.}~\bibnamefont
  {Banerjee}}, \bibinfo {author} {\bibfnamefont {M.}~\bibnamefont {Heiblum}},
  \bibinfo {author} {\bibfnamefont {V.}~\bibnamefont {Umansky}}, \bibinfo
  {author} {\bibfnamefont {D.~E.}\ \bibnamefont {Feldman}}, \bibinfo {author}
  {\bibfnamefont {Y.}~\bibnamefont {Oreg}}, \ and\ \bibinfo {author}
  {\bibfnamefont {A.}~\bibnamefont {Stern}},\ }\bibfield  {title} {\enquote
  {\bibinfo {title} {Observation of half-integer thermal {Hall} conductance},}\
  }\href {\doibase 10.1038/s41586-018-0184-1} {\bibfield  {journal} {\bibinfo
  {journal} {Nature (London)}\ }\textbf {\bibinfo {volume} {559}},\ \bibinfo
  {pages} {205} (\bibinfo {year} {2018})}\BibitemShut {NoStop}%
\bibitem [{\citenamefont {Mross}\ \emph {et~al.}(2018)\citenamefont {Mross},
  \citenamefont {Oreg}, \citenamefont {Stern}, \citenamefont {Margalit},\ and\
  \citenamefont {Heiblum}}]{Mross_theory_2018}%
  \BibitemOpen
  \bibfield  {author} {\bibinfo {author} {\bibfnamefont {D.~F.}\ \bibnamefont
  {Mross}}, \bibinfo {author} {\bibfnamefont {Y.}~\bibnamefont {Oreg}},
  \bibinfo {author} {\bibfnamefont {A.}~\bibnamefont {Stern}}, \bibinfo
  {author} {\bibfnamefont {G.}~\bibnamefont {Margalit}}, \ and\ \bibinfo
  {author} {\bibfnamefont {M.}~\bibnamefont {Heiblum}},\ }\bibfield  {title}
  {\enquote {\bibinfo {title} {Theory of disorder-induced half-integer thermal
  {Hall} conductance},}\ }\href {\doibase 10.1103/PhysRevLett.121.026801}
  {\bibfield  {journal} {\bibinfo  {journal} {Phys. Rev. Lett.}\ }\textbf
  {\bibinfo {volume} {121}},\ \bibinfo {pages} {026801} (\bibinfo {year}
  {2018})}\BibitemShut {NoStop}%
\bibitem [{\citenamefont {Lian}\ and\ \citenamefont
  {Wang}(2018)}]{Lian_theory_2018}%
  \BibitemOpen
  \bibfield  {author} {\bibinfo {author} {\bibfnamefont {B.}~\bibnamefont
  {Lian}}\ and\ \bibinfo {author} {\bibfnamefont {J.}~\bibnamefont {Wang}},\
  }\bibfield  {title} {\enquote {\bibinfo {title} {Theory of the disordered
  $\nu=$~5/2 quantum thermal {Hall} state: Emergent symmetry and phase
  diagram},}\ }\href {\doibase 10.1103/PhysRevB.97.165124} {\bibfield
  {journal} {\bibinfo  {journal} {Phys. Rev. B}\ }\textbf {\bibinfo {volume}
  {97}},\ \bibinfo {pages} {165124} (\bibinfo {year} {2018})}\BibitemShut
  {NoStop}%
\bibitem [{\citenamefont {Wang}\ \emph {et~al.}(2018)\citenamefont {Wang},
  \citenamefont {Vishwanath},\ and\ \citenamefont
  {Halperin}}]{Wang_topological_2018}%
  \BibitemOpen
  \bibfield  {author} {\bibinfo {author} {\bibfnamefont {C.}~\bibnamefont
  {Wang}}, \bibinfo {author} {\bibfnamefont {A.}~\bibnamefont {Vishwanath}}, \
  and\ \bibinfo {author} {\bibfnamefont {B.~I.}\ \bibnamefont {Halperin}},\
  }\bibfield  {title} {\enquote {\bibinfo {title} {Topological order from
  disorder and the quantized {Hall} thermal metal: Possible applications to the
  $\nu=5/2$ state},}\ }\href {\doibase 10.1103/PhysRevB.98.045112} {\bibfield
  {journal} {\bibinfo  {journal} {Phys. Rev. B}\ }\textbf {\bibinfo {volume}
  {98}},\ \bibinfo {pages} {045112} (\bibinfo {year} {2018})}\BibitemShut
  {NoStop}%
\bibitem [{\citenamefont
  {Simon}(2018{\natexlab{a}})}]{Simon_equilibration_2018}%
  \BibitemOpen
  \bibfield  {author} {\bibinfo {author} {\bibfnamefont {S.~H.}\ \bibnamefont
  {Simon}},\ }\bibfield  {title} {\enquote {\bibinfo {title} {Interpretation of
  thermal conductance of the $\nu=$~5/2 edge},}\ }\href {\doibase
  10.1103/PhysRevB.97.121406} {\bibfield  {journal} {\bibinfo  {journal} {Phys.
  Rev. B}\ }\textbf {\bibinfo {volume} {97}},\ \bibinfo {pages} {121406(R)}
  (\bibinfo {year} {2018}{\natexlab{a}})}\BibitemShut {NoStop}%
\bibitem [{\citenamefont {Feldman}(2018)}]{Feldman_comment_2018}%
  \BibitemOpen
  \bibfield  {author} {\bibinfo {author} {\bibfnamefont {D.~E.}\ \bibnamefont
  {Feldman}},\ }\bibfield  {title} {\enquote {\bibinfo {title} {Comment on
  ``interpretation of thermal conductance of the $\nu=$~5/2 edge''},}\ }\href
  {\doibase 10.1103/PhysRevB.98.167401} {\bibfield  {journal} {\bibinfo
  {journal} {Phys. Rev. B}\ }\textbf {\bibinfo {volume} {98}},\ \bibinfo
  {pages} {167401} (\bibinfo {year} {2018})}\BibitemShut {NoStop}%
\bibitem [{\citenamefont
  {Simon}(2018{\natexlab{b}})}]{Simon_reply_comment_2018}%
  \BibitemOpen
  \bibfield  {author} {\bibinfo {author} {\bibfnamefont {S.~H.}\ \bibnamefont
  {Simon}},\ }\bibfield  {title} {\enquote {\bibinfo {title} {Reply to
  ``{C}omment on `{Interpretation} of thermal conductance of the $\nu=$~5/2
  edge' ''},}\ }\href {\doibase 10.1103/PhysRevB.98.167402} {\bibfield
  {journal} {\bibinfo  {journal} {Phys. Rev. B}\ }\textbf {\bibinfo {volume}
  {98}},\ \bibinfo {pages} {167402} (\bibinfo {year}
  {2018}{\natexlab{b}})}\BibitemShut {NoStop}%
\bibitem [{\citenamefont {Ma}\ and\ \citenamefont
  {Feldman}(2019)}]{Feldman_equilibration_2019}%
  \BibitemOpen
  \bibfield  {author} {\bibinfo {author} {\bibfnamefont {K.~K.~W.}\
  \bibnamefont {Ma}}\ and\ \bibinfo {author} {\bibfnamefont {D.~E.}\
  \bibnamefont {Feldman}},\ }\bibfield  {title} {\enquote {\bibinfo {title}
  {Partial equilibration of integer and fractional edge channels in the thermal
  quantum {Hall} effect},}\ }\href {\doibase 10.1103/PhysRevB.99.085309}
  {\bibfield  {journal} {\bibinfo  {journal} {Phys. Rev. B}\ }\textbf {\bibinfo
  {volume} {99}},\ \bibinfo {pages} {085309} (\bibinfo {year}
  {2019})}\BibitemShut {NoStop}%
\bibitem [{\citenamefont {Simon}\ and\ \citenamefont
  {Rosenow}(2020)}]{Simon_equilibration_2020}%
  \BibitemOpen
  \bibfield  {author} {\bibinfo {author} {\bibfnamefont {S.~H.}\ \bibnamefont
  {Simon}}\ and\ \bibinfo {author} {\bibfnamefont {B.}~\bibnamefont
  {Rosenow}},\ }\bibfield  {title} {\enquote {\bibinfo {title} {Partial
  equilibration of the anti-{Pfaffian} edge due to {Majorana} disorder},}\
  }\href {\doibase 10.1103/PhysRevLett.124.126801} {\bibfield  {journal}
  {\bibinfo  {journal} {Phys. Rev. Lett.}\ }\textbf {\bibinfo {volume} {124}},\
  \bibinfo {pages} {126801} (\bibinfo {year} {2020})}\BibitemShut {NoStop}%
\bibitem [{\citenamefont {Asasi}\ and\ \citenamefont
  {Mulligan}(2020)}]{Asasi_equilibration_2020}%
  \BibitemOpen
  \bibfield  {author} {\bibinfo {author} {\bibfnamefont {Hamed}\ \bibnamefont
  {Asasi}}\ and\ \bibinfo {author} {\bibfnamefont {Michael}\ \bibnamefont
  {Mulligan}},\ }\bibfield  {title} {\enquote {\bibinfo {title} {Partial
  equilibration of anti-pfaffian edge modes at $\ensuremath{\nu}=5/2$},}\
  }\href {\doibase 10.1103/PhysRevB.102.205104} {\bibfield  {journal} {\bibinfo
   {journal} {Phys. Rev. B}\ }\textbf {\bibinfo {volume} {102}},\ \bibinfo
  {pages} {205104} (\bibinfo {year} {2020})}\BibitemShut {NoStop}%
\bibitem [{\citenamefont {Zucker}\ and\ \citenamefont
  {Feldman}(2016)}]{Zucker_stabilization_2016}%
  \BibitemOpen
  \bibfield  {author} {\bibinfo {author} {\bibfnamefont {P.~T.}\ \bibnamefont
  {Zucker}}\ and\ \bibinfo {author} {\bibfnamefont {D.~E.}\ \bibnamefont
  {Feldman}},\ }\bibfield  {title} {\enquote {\bibinfo {title} {Stabilization
  of the particle-hole {Pfaffian} order by {Landau}-level mixing and impurities
  that break particle-hole symmetry},}\ }\href {\doibase
  10.1103/PhysRevLett.117.096802} {\bibfield  {journal} {\bibinfo  {journal}
  {Phys. Rev. Lett.}\ }\textbf {\bibinfo {volume} {117}},\ \bibinfo {pages}
  {096802} (\bibinfo {year} {2016})}\BibitemShut {NoStop}%
\bibitem [{\citenamefont {Yang}(2017)}]{Yang_particle-hole_2017}%
  \BibitemOpen
  \bibfield  {author} {\bibinfo {author} {\bibfnamefont {J.}~\bibnamefont
  {Yang}},\ }\bibfield  {title} {\enquote {\bibinfo {title} {Particle-hole
  symmetry and the fractional quantum {Hall} states at $5/2$ filling factor},}\
  }\href {http://arxiv.org/abs/1701.03562} {\bibfield  {journal} {\bibinfo
  {journal} {arXiv:1701.03562}\ } (\bibinfo {year} {2017})}\BibitemShut
  {NoStop}%
\bibitem [{\citenamefont {Jolicoeur}(2007)}]{Jolicoeur_new_series_2007}%
  \BibitemOpen
  \bibfield  {author} {\bibinfo {author} {\bibfnamefont {Th.}\ \bibnamefont
  {Jolicoeur}},\ }\bibfield  {title} {\enquote {\bibinfo {title} {Non-abelian
  states with negative flux: A new series of quantum hall states},}\ }\href
  {\doibase 10.1103/PhysRevLett.99.036805} {\bibfield  {journal} {\bibinfo
  {journal} {Phys. Rev. Lett.}\ }\textbf {\bibinfo {volume} {99}},\ \bibinfo
  {pages} {036805} (\bibinfo {year} {2007})}\BibitemShut {NoStop}%
\bibitem [{\citenamefont {Balram}\ \emph {et~al.}(2018)\citenamefont {Balram},
  \citenamefont {Barkeshli},\ and\ \citenamefont
  {Rudner}}]{Balram_parton_2018}%
  \BibitemOpen
  \bibfield  {author} {\bibinfo {author} {\bibfnamefont {A.~C.}\ \bibnamefont
  {Balram}}, \bibinfo {author} {\bibfnamefont {M.}~\bibnamefont {Barkeshli}}, \
  and\ \bibinfo {author} {\bibfnamefont {M.~S.}\ \bibnamefont {Rudner}},\
  }\bibfield  {title} {\enquote {\bibinfo {title} {Parton construction of a
  wave function in the anti-{Pfaffian} phase},}\ }\href {\doibase
  10.1103/PhysRevB.98.035127} {\bibfield  {journal} {\bibinfo  {journal} {Phys.
  Rev. B}\ }\textbf {\bibinfo {volume} {98}},\ \bibinfo {pages} {035127}
  (\bibinfo {year} {2018})}\BibitemShut {NoStop}%
\bibitem [{\citenamefont {{Mishmash}}\ \emph {et~al.}(2018)\citenamefont
  {{Mishmash}}, \citenamefont {{Mross}}, \citenamefont {{Alicea}},\ and\
  \citenamefont {{Motrunich}}}]{Mishmash_numerical_2018}%
  \BibitemOpen
  \bibfield  {author} {\bibinfo {author} {\bibfnamefont {R.~V.}\ \bibnamefont
  {{Mishmash}}}, \bibinfo {author} {\bibfnamefont {D.~F.}\ \bibnamefont
  {{Mross}}}, \bibinfo {author} {\bibfnamefont {J.}~\bibnamefont {{Alicea}}}, \
  and\ \bibinfo {author} {\bibfnamefont {O.~I.}\ \bibnamefont {{Motrunich}}},\
  }\bibfield  {title} {\enquote {\bibinfo {title} {{Numerical exploration of
  trial wave functions for the particle-hole-symmetric Pfaffian}},}\ }\href
  {\doibase 10.1103/PhysRevB.98.081107} {\bibfield  {journal} {\bibinfo
  {journal} {Phys. Rev. B}\ }\textbf {\bibinfo {volume} {98}},\ \bibinfo
  {pages} {081107(R)} (\bibinfo {year} {2018})}\BibitemShut {NoStop}%
\bibitem [{\citenamefont {M\"oller}\ and\ \citenamefont
  {Simon}(2008)}]{moller_paired_2008}%
  \BibitemOpen
  \bibfield  {author} {\bibinfo {author} {\bibfnamefont {G.}~\bibnamefont
  {M\"oller}}\ and\ \bibinfo {author} {\bibfnamefont {S.~H.}\ \bibnamefont
  {Simon}},\ }\bibfield  {title} {\enquote {\bibinfo {title} {Paired
  composite-fermion wave functions},}\ }\href {\doibase
  10.1103/PhysRevB.77.075319} {\bibfield  {journal} {\bibinfo  {journal} {Phys.
  Rev. B}\ }\textbf {\bibinfo {volume} {77}},\ \bibinfo {pages} {075319}
  (\bibinfo {year} {2008})}\BibitemShut {NoStop}%
\bibitem [{\citenamefont {M{\"o}ller}(2006)}]{Moller_phd}%
  \BibitemOpen
  \bibfield  {author} {\bibinfo {author} {\bibfnamefont {Gunnar}\ \bibnamefont
  {M{\"o}ller}},\ }\emph {\bibinfo {title} {{Dynamically reduced spaces in
  condensed matter physics:<br />Quantum Hall bilayers, dimensional reduction,
  and magnetic spin systems}}},\ \href
  {https://tel.archives-ouvertes.fr/tel-00121765} {\bibinfo {type} {Theses}},\
  \bibinfo  {school} {{Universit{\'e} Paris Sud - Paris XI}} (\bibinfo {year}
  {2006})\BibitemShut {NoStop}%
\bibitem [{\citenamefont {Wen}\ and\ \citenamefont
  {Zee}(1992)}]{Wen_shift_1992}%
  \BibitemOpen
  \bibfield  {author} {\bibinfo {author} {\bibfnamefont {X.~G.}\ \bibnamefont
  {Wen}}\ and\ \bibinfo {author} {\bibfnamefont {A.}~\bibnamefont {Zee}},\
  }\bibfield  {title} {\enquote {\bibinfo {title} {Shift and spin vector: {New}
  topological quantum numbers for the {Hall} fluids},}\ }\href {\doibase
  10.1103/PhysRevLett.69.953} {\bibfield  {journal} {\bibinfo  {journal} {Phys.
  Rev. Lett.}\ }\textbf {\bibinfo {volume} {69}},\ \bibinfo {pages} {953}
  (\bibinfo {year} {1992})}\BibitemShut {NoStop}%
\bibitem [{Note1()}]{Note1}%
  \BibitemOpen
  \bibinfo {note} {Monopole harmonics and their properties were described in
  Refs.~\protect \rev@citealpnum {Wu_dirac_1976,Wu_properties_1977}; we use the
  conventions of Ref.~\protect \rev@citealpnum
  {Jain_composite_2007}}\BibitemShut {NoStop}%
\bibitem [{\citenamefont {M\"oller}\ \emph {et~al.}(2009)\citenamefont
  {M\"oller}, \citenamefont {Simon},\ and\ \citenamefont
  {Rezayi}}]{Moller_Trial_2009}%
  \BibitemOpen
  \bibfield  {author} {\bibinfo {author} {\bibfnamefont {Gunnar}\ \bibnamefont
  {M\"oller}}, \bibinfo {author} {\bibfnamefont {Steven~H.}\ \bibnamefont
  {Simon}}, \ and\ \bibinfo {author} {\bibfnamefont {Edward~H.}\ \bibnamefont
  {Rezayi}},\ }\bibfield  {title} {\enquote {\bibinfo {title} {Trial wave
  functions for $\ensuremath{\nu}=\frac{1}{2}+\frac{1}{2}$ quantum hall
  bilayers},}\ }\href {\doibase 10.1103/PhysRevB.79.125106} {\bibfield
  {journal} {\bibinfo  {journal} {Phys. Rev. B}\ }\textbf {\bibinfo {volume}
  {79}},\ \bibinfo {pages} {125106} (\bibinfo {year} {2009})}\BibitemShut
  {NoStop}%
\bibitem [{\citenamefont {Wang}\ \emph {et~al.}(2019)\citenamefont {Wang},
  \citenamefont {Geraedts}, \citenamefont {Rezayi},\ and\ \citenamefont
  {Haldane}}]{Wang_MC_PH_conjugation_2019}%
  \BibitemOpen
  \bibfield  {author} {\bibinfo {author} {\bibfnamefont {J.}~\bibnamefont
  {Wang}}, \bibinfo {author} {\bibfnamefont {S.~D.}\ \bibnamefont {Geraedts}},
  \bibinfo {author} {\bibfnamefont {E.~H.}\ \bibnamefont {Rezayi}}, \ and\
  \bibinfo {author} {\bibfnamefont {F.~D.~M.}\ \bibnamefont {Haldane}},\
  }\bibfield  {title} {\enquote {\bibinfo {title} {Lattice {Monte} {Carlo} for
  quantum {Hall} states on a torus},}\ }\href {\doibase
  10.1103/PhysRevB.99.125123} {\bibfield  {journal} {\bibinfo  {journal} {Phys.
  Rev. B}\ }\textbf {\bibinfo {volume} {99}},\ \bibinfo {pages} {125123}
  (\bibinfo {year} {2019})}\BibitemShut {NoStop}%
\bibitem [{\citenamefont {{Davenport}}\ and\ \citenamefont
  {{Simon}}(2012)}]{Davenport_projection_2012}%
  \BibitemOpen
  \bibfield  {author} {\bibinfo {author} {\bibfnamefont {S.~C.}\ \bibnamefont
  {{Davenport}}}\ and\ \bibinfo {author} {\bibfnamefont {S.~H.}\ \bibnamefont
  {{Simon}}},\ }\bibfield  {title} {\enquote {\bibinfo {title} {{Spinful
  composite fermions in a negative effective field}},}\ }\href {\doibase
  10.1103/PhysRevB.85.245303} {\bibfield  {journal} {\bibinfo  {journal} {Phys.
  Rev. B}\ }\textbf {\bibinfo {volume} {85}},\ \bibinfo {eid} {245303}
  (\bibinfo {year} {2012})}\BibitemShut {NoStop}%
\bibitem [{\citenamefont {{Fulsebakke}}(2016)}]{Fulsebakke_projection_2016}%
  \BibitemOpen
  \bibfield  {author} {\bibinfo {author} {\bibfnamefont {J.}~\bibnamefont
  {{Fulsebakke}}},\ }\bibfield  {title} {\enquote {\bibinfo {title}
  {{Projections and correlations in the fractional quantum {Hall} effect}},}\
  }\href {http://mural.maynoothuniversity.ie/7590/} {\bibfield  {journal}
  {\bibinfo  {journal} {Ph.D. thesis}\ } (\bibinfo {year} {2016})}\BibitemShut
  {NoStop}%
\bibitem [{\citenamefont {{Mukherjee}}\ and\ \citenamefont
  {{Mandal}}(2015)}]{Mukherjee_incompressible_2015}%
  \BibitemOpen
  \bibfield  {author} {\bibinfo {author} {\bibfnamefont {S.}~\bibnamefont
  {{Mukherjee}}}\ and\ \bibinfo {author} {\bibfnamefont {S.~S.}\ \bibnamefont
  {{Mandal}}},\ }\bibfield  {title} {\enquote {\bibinfo {title}
  {{Incompressible states of the interacting composite fermions in negative
  effective magnetic fields at $\nu~=$~4/13,~5/17, and 3/10}},}\ }\href
  {\doibase 10.1103/PhysRevB.92.235302} {\bibfield  {journal} {\bibinfo
  {journal} {Phys. Rev. B}\ }\textbf {\bibinfo {volume} {92}},\ \bibinfo {eid}
  {235302} (\bibinfo {year} {2015})}\BibitemShut {NoStop}%
\bibitem [{Note2()}]{Note2}%
  \BibitemOpen
  \bibinfo {note} {Upon rewriting Eq.~\protect \textup {\hbox {\mathsurround
  \z@ \protect \normalfont (\ignorespaces \ref {eqn.addition}\unskip
  \@@italiccorr )}} in the form of Eq.~\protect \textup {\hbox {\mathsurround
  \z@ \protect \normalfont (\ignorespaces \ref {eqn.A_expanssion1}\unskip
  \@@italiccorr )}}, it becomes manifest that the pairing channel is unaffected
  by truncating the infinite sum.}\BibitemShut {Stop}%
\bibitem [{foo()}]{footnote_1}%
  \BibitemOpen
  \href@noop {} {\bibinfo  {journal} {The structure factor is related to the
  real-space correlation function via
  $g(R)=1+\frac{4\pi}{N_p}\sum_{l}\frac{2l+1}{4\pi}(S_l-1) P_{l}[\cos(R)]$. In
  spherically symmetric cases, our definition coincides with that of
  Ref.~\onlinecite{Kamilla_static_1997}}\ }\BibitemShut {NoStop}%
\bibitem [{\citenamefont {Altshuler}\ \emph {et~al.}(1994)\citenamefont
  {Altshuler}, \citenamefont {Ioffe},\ and\ \citenamefont
  {Millis}}]{Altshuler_low_energy_1994}%
  \BibitemOpen
\bibfield  {journal} {  }\bibfield  {author} {\bibinfo {author} {\bibfnamefont
  {B.~L.}\ \bibnamefont {Altshuler}}, \bibinfo {author} {\bibfnamefont {L.~B.}\
  \bibnamefont {Ioffe}}, \ and\ \bibinfo {author} {\bibfnamefont {A.~J.}\
  \bibnamefont {Millis}},\ }\bibfield  {title} {\enquote {\bibinfo {title}
  {Low-energy properties of fermions with singular interactions},}\ }\href
  {\doibase 10.1103/PhysRevB.50.14048} {\bibfield  {journal} {\bibinfo
  {journal} {Phys. Rev. B}\ }\textbf {\bibinfo {volume} {50}},\ \bibinfo
  {pages} {14048} (\bibinfo {year} {1994})}\BibitemShut {NoStop}%
\bibitem [{\citenamefont {Mross}\ \emph {et~al.}(2010)\citenamefont {Mross},
  \citenamefont {McGreevy}, \citenamefont {Liu},\ and\ \citenamefont
  {Senthil}}]{Mross_controlled_expansion_2010}%
  \BibitemOpen
  \bibfield  {author} {\bibinfo {author} {\bibfnamefont {D.~F.}\ \bibnamefont
  {Mross}}, \bibinfo {author} {\bibfnamefont {J.}~\bibnamefont {McGreevy}},
  \bibinfo {author} {\bibfnamefont {H.}~\bibnamefont {Liu}}, \ and\ \bibinfo
  {author} {\bibfnamefont {T.}~\bibnamefont {Senthil}},\ }\bibfield  {title}
  {\enquote {\bibinfo {title} {Controlled expansion for certain
  non-{Fermi}-liquid metals},}\ }\href {\doibase 10.1103/PhysRevB.82.045121}
  {\bibfield  {journal} {\bibinfo  {journal} {Phys. Rev. B}\ }\textbf {\bibinfo
  {volume} {82}},\ \bibinfo {pages} {045121} (\bibinfo {year}
  {2010})}\BibitemShut {NoStop}%
\bibitem [{\citenamefont {Sheng}\ \emph {et~al.}(2009)\citenamefont {Sheng},
  \citenamefont {Motrunich},\ and\ \citenamefont
  {Fisher}}]{Sheng_spin_bose_2009}%
  \BibitemOpen
  \bibfield  {author} {\bibinfo {author} {\bibfnamefont {D.~N.}\ \bibnamefont
  {Sheng}}, \bibinfo {author} {\bibfnamefont {O.~I.}\ \bibnamefont
  {Motrunich}}, \ and\ \bibinfo {author} {\bibfnamefont {M.~P.~A.}\
  \bibnamefont {Fisher}},\ }\bibfield  {title} {\enquote {\bibinfo {title}
  {Spin {Bose}-metal phase in a spin-$\frac{1}{2}$ model with ring exchange on
  a two-leg triangular strip},}\ }\href {\doibase 10.1103/PhysRevB.79.205112}
  {\bibfield  {journal} {\bibinfo  {journal} {Phys. Rev. B}\ }\textbf {\bibinfo
  {volume} {79}},\ \bibinfo {pages} {205112} (\bibinfo {year}
  {2009})}\BibitemShut {NoStop}%
\bibitem [{\citenamefont {M\"oller}\ \emph {et~al.}(2011)\citenamefont
  {M\"oller}, \citenamefont {W\'ojs},\ and\ \citenamefont
  {Cooper}}]{Moller_Neutral_2011}%
  \BibitemOpen
  \bibfield  {author} {\bibinfo {author} {\bibfnamefont {Gunnar}\ \bibnamefont
  {M\"oller}}, \bibinfo {author} {\bibfnamefont {Arkadiusz}\ \bibnamefont
  {W\'ojs}}, \ and\ \bibinfo {author} {\bibfnamefont {Nigel~R.}\ \bibnamefont
  {Cooper}},\ }\bibfield  {title} {\enquote {\bibinfo {title} {Neutral fermion
  excitations in the moore-read state at filling factor
  $\ensuremath{\nu}=5/2$},}\ }\href {\doibase 10.1103/PhysRevLett.107.036803}
  {\bibfield  {journal} {\bibinfo  {journal} {Phys. Rev. Lett.}\ }\textbf
  {\bibinfo {volume} {107}},\ \bibinfo {pages} {036803} (\bibinfo {year}
  {2011})}\BibitemShut {NoStop}%
\bibitem [{\citenamefont {Bonderson}\ \emph {et~al.}(2011)\citenamefont
  {Bonderson}, \citenamefont {Feiguin},\ and\ \citenamefont
  {Nayak}}]{Bonderson_num_corrlength_2011}%
  \BibitemOpen
  \bibfield  {author} {\bibinfo {author} {\bibfnamefont {P.}~\bibnamefont
  {Bonderson}}, \bibinfo {author} {\bibfnamefont {A.~E.}\ \bibnamefont
  {Feiguin}}, \ and\ \bibinfo {author} {\bibfnamefont {C.}~\bibnamefont
  {Nayak}},\ }\bibfield  {title} {\enquote {\bibinfo {title} {Numerical
  calculation of the neutral fermion gap at the $\ensuremath{\nu}=5/2$
  fractional quantum {Hall} state},}\ }\href {\doibase
  10.1103/PhysRevLett.106.186802} {\bibfield  {journal} {\bibinfo  {journal}
  {Phys. Rev. Lett.}\ }\textbf {\bibinfo {volume} {106}},\ \bibinfo {pages}
  {186802} (\bibinfo {year} {2011})}\BibitemShut {NoStop}%
\bibitem [{\citenamefont {Baraban}\ \emph {et~al.}(2009)\citenamefont
  {Baraban}, \citenamefont {Zikos}, \citenamefont {Bonesteel},\ and\
  \citenamefont {Simon}}]{Baraban_num_corrlength_2009}%
  \BibitemOpen
  \bibfield  {author} {\bibinfo {author} {\bibfnamefont {M.}~\bibnamefont
  {Baraban}}, \bibinfo {author} {\bibfnamefont {G.}~\bibnamefont {Zikos}},
  \bibinfo {author} {\bibfnamefont {N.}~\bibnamefont {Bonesteel}}, \ and\
  \bibinfo {author} {\bibfnamefont {S.~H.}\ \bibnamefont {Simon}},\ }\bibfield
  {title} {\enquote {\bibinfo {title} {Numerical analysis of quasiholes of the
  {Moore}-{Read} wave function},}\ }\href {\doibase
  10.1103/PhysRevLett.103.076801} {\bibfield  {journal} {\bibinfo  {journal}
  {Phys. Rev. Lett.}\ }\textbf {\bibinfo {volume} {103}},\ \bibinfo {pages}
  {076801} (\bibinfo {year} {2009})}\BibitemShut {NoStop}%
\bibitem [{\citenamefont {Kamilla}\ \emph {et~al.}(1997)\citenamefont
  {Kamilla}, \citenamefont {Jain},\ and\ \citenamefont
  {Girvin}}]{Kamilla_static_1997}%
  \BibitemOpen
  \bibfield  {author} {\bibinfo {author} {\bibfnamefont {R.~K.}\ \bibnamefont
  {Kamilla}}, \bibinfo {author} {\bibfnamefont {J.~K.}\ \bibnamefont {Jain}}, \
  and\ \bibinfo {author} {\bibfnamefont {S.~M.}\ \bibnamefont {Girvin}},\
  }\bibfield  {title} {\enquote {\bibinfo {title} {Fermi-sea-like correlations
  in a partially filled {Landau} level},}\ }\href {\doibase
  10.1103/PhysRevB.56.12411} {\bibfield  {journal} {\bibinfo  {journal} {Phys.
  Rev. B}\ }\textbf {\bibinfo {volume} {56}},\ \bibinfo {pages} {12411}
  (\bibinfo {year} {1997})}\BibitemShut {NoStop}%
\bibitem [{\citenamefont {Wu}\ and\ \citenamefont
  {Yang}(1976)}]{Wu_dirac_1976}%
  \BibitemOpen
  \bibfield  {author} {\bibinfo {author} {\bibfnamefont {T.~T.}\ \bibnamefont
  {Wu}}\ and\ \bibinfo {author} {\bibfnamefont {C.~N.}\ \bibnamefont {Yang}},\
  }\bibfield  {title} {\enquote {\bibinfo {title} {Dirac monopole without
  strings: {Monopole} harmonics},}\ }\href {\doibase
  http://dx.doi.org/10.1016/0550-3213(76)90143-7} {\bibfield  {journal}
  {\bibinfo  {journal} {Nucl. Phys. B}\ }\textbf {\bibinfo {volume} {107}},\
  \bibinfo {pages} {365} (\bibinfo {year} {1976})}\BibitemShut {NoStop}%
\bibitem [{\citenamefont {Wu}\ and\ \citenamefont
  {Yang}(1977)}]{Wu_properties_1977}%
  \BibitemOpen
  \bibfield  {author} {\bibinfo {author} {\bibfnamefont {T.~T.}\ \bibnamefont
  {Wu}}\ and\ \bibinfo {author} {\bibfnamefont {C.~N.}\ \bibnamefont {Yang}},\
  }\bibfield  {title} {\enquote {\bibinfo {title} {Some properties of monopole
  harmonics},}\ }\href {\doibase 10.1103/PhysRevD.16.1018} {\bibfield
  {journal} {\bibinfo  {journal} {Phys. Rev. D}\ }\textbf {\bibinfo {volume}
  {16}},\ \bibinfo {pages} {1018} (\bibinfo {year} {1977})}\BibitemShut
  {NoStop}%
\bibitem [{Note3()}]{Note3}%
  \BibitemOpen
  \bibinfo {note} {One can always find a conformal transformation that maps
  $\protect \bm {r}_a$ and $\protect \bm {r}_b$ to opposite poles and thus $h_i
  \rightarrow u_i/v_i$, while the cross-ratio $\omega _{ab}\Omega _{ij}$ is
  conformally invariant.}\BibitemShut {Stop}%
\bibitem [{\citenamefont {Wimmer}(2012)}]{pfapack_Wimmer_2012}%
  \BibitemOpen
  \bibfield  {author} {\bibinfo {author} {\bibfnamefont {M.}~\bibnamefont
  {Wimmer}},\ }\bibfield  {title} {\enquote {\bibinfo {title} {Algorithm 923:
  Efficient numerical computation of the {Pfaffian} for dense and banded
  skew-symmetric matrices},}\ }\href {\doibase 10.1145/2331130.2331138}
  {\bibfield  {journal} {\bibinfo  {journal} {ACM Trans. Math. Softw.}\
  }\textbf {\bibinfo {volume} {38}} (\bibinfo {year} {2012}),\
  10.1145/2331130.2331138}\BibitemShut {NoStop}%
\bibitem [{\citenamefont {Li}\ and\ \citenamefont
  {Haldane}(2008)}]{Li_entanglement_spectrum_2008}%
  \BibitemOpen
  \bibfield  {author} {\bibinfo {author} {\bibfnamefont {H.}~\bibnamefont
  {Li}}\ and\ \bibinfo {author} {\bibfnamefont {F.~D.~M.}\ \bibnamefont
  {Haldane}},\ }\bibfield  {title} {\enquote {\bibinfo {title} {Entanglement
  spectrum as a generalization of entanglement entropy: Identification of
  topological order in non-{Abelian} fractional quantum {Hall} effect
  states},}\ }\href {\doibase 10.1103/PhysRevLett.101.010504} {\bibfield
  {journal} {\bibinfo  {journal} {Phys. Rev. Lett.}\ }\textbf {\bibinfo
  {volume} {101}},\ \bibinfo {pages} {010504} (\bibinfo {year}
  {2008})}\BibitemShut {NoStop}%
\end{thebibliography}%

\end{document}